\documentclass[twocolumn,preprintnumbers,amsmath,amssymb,floatfix]{revtex4}
\usepackage{epsfig}
\usepackage{amsmath}
\graphicspath{{figures/}}
\usepackage{epsfig}
\usepackage{amsmath}
\usepackage{color}
\usepackage{graphicx,epsfig}
\usepackage{color}
\usepackage[utf8]{inputenc}
\usepackage[T1]{fontenc}
\usepackage{float}
\usepackage{natbib}
\usepackage{caption}
\usepackage{subcaption}
\usepackage[newcommands]{ragged2e}
\usepackage{caption}
\begin{document}

\title{Three-phase equilibria of hydrates from computer simulation. \\I. Finite-size effects in the methane hydrate}


\author{S. Blazquez$^1$}
\author{J. Algaba$^2$}
\author{J. M. Míguez$^2$}
\author{C. Vega$^1$}
\author{F. J. Blas$^{2*}$}
\author{M. M. Conde$^{3*}$}

\affiliation{$^1$Departamento de Qu\'{\i}mica F\'{\i}sica, Facultad de Ciencias Qu\'{\i}micas, Universidad Complutense de Madrid, 28040 Madrid, Spain}

\affiliation{$^2$Laboratorio de Simulación Molecular y Química Computacional, CIQSO-Centro de Investigación en Química Sostenible and
Departamento de Ciencias Integradas, Universidad de Huelva, 21006 Huelva, Spain
}

\affiliation{$^3$Departamento de Ingeniería Química Industrial y del Medio Ambiente, Escuela Técnica Superior de Ingenieros Industriales, Universidad Politécnica de Madrid,
28006, Madrid, Spain.
}

%

\begin{abstract}

Clathrate hydrates are vital in energy research and environmental applications. Understanding their stability is crucial for harnessing their potential. In this work, we employ direct coexistence simulations to study finite-size effects in the determination of the three-phase equilibrium temperature ($T_3$) for methane hydrates. Two popular water models, TIP4P/Ice and TIP4P/2005, are employed, exploring various system sizes by varying the number of molecules in the hydrate, liquid, and gas phases. The results reveal that finite-size effects play a crucial role in determining $T_3$. The study includes nine configurations with varying system sizes, demonstrating that smaller systems, particularly those leading to stoichiometric conditions and bubble formation, may yield inaccurate $T_3$ values. The emergence of methane bubbles within the liquid phase, observed in smaller configurations, significantly influences the behavior of the system and can lead to erroneous temperature estimations. Our findings reveal finite size effects on the calculation of the $T_3$ by direct coexistence simulations and clarify the system size convergence for both models, shedding light on discrepancies found in the literature. The results contribute to a deeper understanding of the phase equilibrium of gas hydrates and offer valuable information for future research in this field.

\end{abstract}

\maketitle
$^*$Corresponding author: maria.mconde@upm.es and felipe@uhu.es


\section{Introduction}

Clathrate hydrates represent a fascinating class of materials that possess the unique capability to encapsulate small molecules of great interest, such as methane (CH$_4$) carbon dioxide (CO$_2$) or even cyclopentane, within their crystalline lattice formed by water molecules \cite{Sloan_book_hydrates,PTRSL_1811_101_1,Nature_1987_325_135,mcmullan1965polyhedral,mak1965polyhedral,mallek2022experimental,mallek2022investigation}. Their importance is underscored by their diverse applications in energy storage, with a particular emphasis on the exploration of hydrogen hydrates as a promising avenue \cite{N_434_743_2005,JPCB_113_7558_2009,N_2003_426_353,S_2004_306_469,JPCB_112_1885_2008,JCP_130_014506_2009}. Additionally, gas hydrates offer a potential solution for mitigating the emission of CO$_2$, a prominent greenhouse gas, by serving as a means for its capture \cite{Aya-CO2,IJT_17_1_1996,Herzog-CO2,yezdimer}. Beyond their presence on Earth, they also play a significant role in icy satellites within our solar system under conditions of salty water, thereby influencing the formation conditions \cite{JPCC_1_80_2020,https://doi.org/10.1002/2014RG000463,PCCP_2017_19_9566,D1CP02638K,PRIETOBALLESTEROS2005491,KARGEL2000226,10.1130/G22311.1}.

However, among the various hydrates, none have garnered as much attention as methane hydrates. Methane, the primary constituent of natural gas, is one of the most widely used energy sources. 
In addition to known natural gas reserves in terrestrial regions, vast deposits of natural gas in hydrate form are found in marine sediment beds \cite{Sloan_book_hydrates,CG_1988_71_41,ARE_1990_15_53,GRL_34_L22303_2007,N_2007_445_303,N_2003_426_353}. Recently, the first controlled extractions of methane from oceanic floors have begun \cite{YU20195213,C9RA00755E,10.4043/25243-MS,doi:10.1021/acs.energyfuels.6b03143}, opening a new pathway for the use of natural gas as an energy source. The thermodynamic conditions necessary for hydrate stability involve low temperatures and high pressures, making the knowledge and control of stability conditions particularly vital. Furthermore, recent findings by Ketzer \textit{et al.} \cite{ketzer2020gas} have highlighted the potential environmental consequences of global ocean warming caused by the release of methane stored in hydrate form from the seabed.

In recent years, the use of computer simulations in the study of gas hydrates has become a highly valuable tool
\cite{JCP_2010_133_064507,noya2011path,krishnan2022controlling,JCP_149_074502_2018,doi:10.1143/JPSJS.81SA.SA027,JPCB_118_1900_2014,JPCB_118_11797_2014,JPCC_120_3305_2016,JPCC_123_1806_2019,JPCC_1_80_2020,JPCC_122_2673_2018,SUSILO20086,doi:10.1021/jp807208z,doi:10.1021/jp4023772,KONDORI2019264,doi:10.1021/acs.jpclett.8b01210,doi:10.1021/acscentsci.8b00755,doi:10.1021/jp504852k,Walsh1095,doi:10.1063/1.5084785,BRUMBY2016242,JCP_2015_142_124505,doi:10.1021/acs.jpcc.5b05393,doi:10.1139/cjc-2015-0003,KONDORI2017754,doi:10.1021/jp102874s,doi:10.1021/acs.jpcb.8b02285,doi:10.1021/jp5002012,doi:10.1021/jz3012113,doi:10.1063/1.4866143,doi:10.1021/ja309117d,doi:10.1021/jp107269q,doi:10.1021/acs.cgd.0c01303,doi:10.1021/jp507959q,grabowska2022solubility,grabowska2023homogeneous,algaba2023solubility,fang2017modeling,fang2023effects,fang2024microscopic,nakayama2009augmented,hakim2010thermodynamic,cao2020mechanical,katsumasa2007thermodynamic,fernandez2022molecular,perez2017computational,sarupria2011molecular,defever2017nucleation,bagherzadeh2015formation,alavi2005molecular,alavi2006molecular,alavi2007hydrogen}. 
In particular, classical simulation methods rely on empirical models to study gas hydrates. Currently, there are numerous force fields available for simulating both the network of water molecules and the gas molecules trapped within. How can we ensure the suitability of these models and the precision of their predictions? Determining the three-phase coexistence line serves as a robust test for the model. Understanding the formation-dissociation conditions of a specific gas hydrate system in terms of pressure and temperature allows exploration of other phenomena in the region of appropriate stability within the phase diagram.

The determination of the three-phase equilibrium line for a hydrate-like system can be studied through Molecular Dynamics simulations using the direct coexistence technique \cite{ladd1977triple,ladd1978interfacial,cape1978molecular,morris2002melting,karim1988ice,karim1990ice,bryk2002ice,bryk2004ice,wang2005melting,nada2004clear,carignano2005molecular}. In 2010, some of us applied this method to estimate the three-phase equilibrium temperature ($T_3$) at various pressures along the equilibrium line (methane hydrate - liquid water - methane gas), using different water models (TIP4P \cite{tip4p}, TIP4P/2005 \cite{abascal05b}, and TIP4P/Ice \cite{JCP_2005_122_234511}). Basically, we put the hydrate phase into contact with the water phase on one side and the gas phase on the other side, ensuring that, due to periodic boundary conditions, every phase was in contact with the other two phases. At 400 bar, the experimental value of $T_3$ is 297 K \cite{Sloan_book_hydrates}. For the TIP4P/Ice model at 400 bar, we obtained two $T_3$ values, 297(8) K and 302(3) K, depending on the system size, larger and smaller, respectively \cite{JCP_2010_133_064507,conde2013note}. In the same year, Jensen \textit{et al.} \cite{JPCB_2010_114_5775} determined the $T_3$ of methane hydrate at 400 bar using the TIP4P/Ice force field and employing Monte Carlo simulations and thermodynamic integration. They obtained a $T_3$ of 314(7) K, deviating significantly above both the experimental value and our values obtained through direct coexistence. 

Years later, and following the direct coexistence methodology, Michalis \textit{et al.} \cite{JCP_2015_142_044501} designed a ``sandwich'' type setup in which the hydrate phase was surrounded by water on both sides, and the gas phase was in contact with water. Using this setup, they obtained a $T_3$ = 293.4(0.9) K at 400 bar, significantly below our value obtained also by direct coexistence and the data obtained by Jensen in 2010. In 2019, Fernandez-Fernandez \textit{et al.} \cite{JML_2019_274_0426} performed a study on methane hydrates under marine conditions. In their study, they used a traditional three-phase direct coexistence setup and obtained a value for $T_3$ close to that provided by Michales \textit{et al.}. Finally, we have revisited the issue in a recent work \cite{grabowska2022solubility} on the solubility of methane in water, obtaining a new value of $T_3$ = 294(2) K at 400 bar for a large-scale system. This revised value falls within the uncertainty of our result for large systems and is in agreement with the values reported by Michalis \textit{et al.} and Fernandez-Fernandez \textit{et al.}. A compilation of the $T_3$ values published in the literature is provided in Table \ref{tabla-pure}.

All of these results reveal that finite-size effects play an important role and should not be underestimated in the determination of $T_3$. This observation is consistent with previous research where finite-size effects have been studied in other systems, including the determination of critical parameters for Lennard-Jones potentials in nucleation simulations \cite{panagiotopoulos1994molecular} and surface tension calculations \cite{orea2005oscillatory,binder2000computer}. These effects have also been studied in square well fluids \cite{vortler2008simulation}. Furthermore, a recent work has performed a complete analysis of the impact of finite sizes on the determination of the melting temperature of ice \cite{conde2017high}. The message is clear: if one aims to determine the equilibrium temperature using the direct coexistence technique, it is necessary to use a minimum system size that allows the neglect of finite-size effects. Advances in computational capabilities now facilitate simulations in larger systems, yielding exceptionally precise estimates of equilibrium temperatures using the direct coexistence technique. The simulations that were very costly in our previous work published in 2010, even for small systems, required many months of calculation time to obtain a $T_3$ value for methane hydrate systems. Today, these simulations can be extended to larger sizes, allowing exploration of finite-size effects and their impact on determining the equilibrium line in gas hydrate systems.

In this work, we carry out a comprehensive study of the finite-size effects present in the determination of the three-phase equilibrium temperature for methane hydrates. We use the TIP4P/Ice and TIP4P/2005 force fields and explore different system sizes by varying the number of molecules in the hydrate, liquid, and gas phases. The structure of this work is as follows: Section II outlines the methodology and simulation details. Section III presents the results of the three-phase equilibrium temperatures for different system sizes of the methane hydrate, a comparison between models, and the size-dependent phenomena observed. Finally, in Section IV, we present the main conclusions of this study.

\begin{table*}
\caption{\label{tabla-pure}\justifying{
Three-phase coexistence temperature ($T_3$) for methane hydrate derived from computational simulations as reported in the literature. All values were obtained at 400 bar. The first column displays the potential model (TIP4P/Ice or TIP4P/2005) used in each simulation study, except for the first value, which refers to the experimental value. The error bars are given in parentheses. The last column indicates the publication year.
}}
\begin{ruledtabular}
\begin{tabular}{c c c c c c c c c c c c c c c}
Model  &&  $T_3$ (K) && Reference &&  Year  \\
\hline
Experimental & & 297 && Sloan \cite{Sloan_book_hydrates}  && 1990 \\
TIP4P/Ice & &   302(3) && Conde et \textit{al.} \cite{JCP_2010_133_064507} &&  2010 \\
TIP4P/Ice & &  297(8) && Conde et \textit{al.} \cite{JCP_2010_133_064507}  && 2010\\
TIP4P/Ice & &  314(7) && Jensen \textit{et al.} \cite{JPCB_2010_114_5775}  && 2010\\
TIP4P/Ice  && 293.4(0.9) && Michalis et \textit{al.} \cite{JCP_2015_142_044501} && 2015 \\
TIP4P/Ice & &  293.5(5) && Fernandez-Fernandez \textit{et al.} \cite{JML_2019_274_0426} &&  2019 \\
TIP4P/Ice  && 290.5(5) && Fernandez-Fernandez \textit{et al.} \cite{JML_2019_274_0426} &&  2019 \\
TIP4P/Ice & &  294(2) && Grabowska \textit{et al.} \cite{grabowska2022solubility}  && 2022 \\
TIP4P/2005 & &  281(2) && Conde et \textit{al.} \cite{JCP_2010_133_064507} && 2010\\
TIP4P/2005 & &  279(1) && Blazquez et \textit{al.} \cite{BLAZQUEZ2023122031} && 2023 \\
\end{tabular}
\end{ruledtabular}
\end{table*}


\section{Methodology and Simulation Details}

To study the stability of methane hydrate and determine the three-phase coexistence temperature ($T_3$) at a specific pressure, we have employed the methodology introduced in 2010 by Conde and Vega \cite{JCP_2010_133_064507}. Inspired by the works on the direct coexistence of two phases \cite{JCP_2006_124_144506,MP_104_3583_2006}, we incorporated three phases in coexistence within the simulation box, i.e., a central phase surrounded on each side by the other two resulting phases. Under constant pressure, $T_3$ is determined by studying the evolution of potential energy as a function of time. The robustness of this methodology has been previously demonstrated in the study of various gas hydrates such as methane hydrate \cite{JCP_2010_133_064507,conde2013note,grabowska2022solubility}, carbon dioxide hydrate \cite{JCP_2015_142_124505}, and methane hydrate in salty water \cite{JML_2019_274_0426}.

Methane hydrate typically adopts the sI structure, characterized by cubic symmetry and the \textit{Pm$\Bar{3}$n} space group. The unit cell comprises eight cages, formed by water molecules, within which methane molecules are trapped. These cages take the form of two types of polyhedra: six tetradecahedra 5$^{12}$6$^2$ and two dodecahedra 5$^{12}$. In total, the unit cell consists of 46 H$_2$O molecules and 8 CH$_4$ molecules.

To build our unit cell, we employed the crystallographic parameters proposed by Yousuf \textit{et al.} \cite{Appphys-2004-78-925}. Furthermore, the water molecules constituting methane hydrate exhibit proton disorder \cite{pauling_book,JCP_1977_66_4699,AC_10_72_1957}. We have employed the algorithm proposed by Buch \textit{et al.} \cite{JPCB_1998_102_08641} to generate solid configurations of the sI hydrate, satisfying the Bernal-Fowler rules \cite{JCP_1933_01_00515} and having a zero or nearly zero dipole moment. For all configurations studied in this work, regardless of size, the initial configuration is composed of a slab of liquid water surrounded on one side by a solid slab of methane hydrate and on the other side by a slab of methane molecules in phase gas. This arrangement ensures that all three phases coexist if periodic boundary conditions apply. 

The main purpose of this work is to explore the role of finite size effects in determining the $T_3$ of methane hydrate. To achieve this, we estimate the $T_3$ in 9 systems of varying sizes. Table \ref{tabla-moleculas} presents the details, including the number of molecules constituting each phase and the unit cell replication factor for the 9 size-dependent configurations. The simulation box size is variable and depends on the number of molecules in each configuration. All interfaces were oriented perpendicular to the $x$-axis.

After generating the initial configurations, direct coexistence simulations are performed in the $NpT$ ensemble at a fixed pressure of 400 bar at different temperatures for each of the 9 configurations. If the simulated temperature falls below $T_3$, the hydrate phase grows, resulting in a decrease in potential energy. Successful hydrate growth requires methane molecules from the gas phase to diffuse through the liquid phase, approaching the hydrate-liquid interface where the typical cages of the sI structure will form, capturing the methane molecules. Conversely, when the simulation temperature is above $T_3$, the hydrate phase melts, resulting in an increase in potential energy. Therefore, $T_3$ can be estimated as the temperature located between the lowest temperature at which the hydrate melts and the highest temperature at which the hydrate grows.

For all direct coexistence simulations of this work, we used the molecular dynamics package GROMACS (version 4.6.5) \cite{spoel05,hess08}.
The leap-frog integration algorithm \cite{bee:jcp76}  was employed with a time step of 2 fs. Periodic boundary conditions were applied in all directions. The temperature was kept constant using the Nos\'e-Hoover thermostat \cite{nose84,hoover85} with a coupling constant of 2 ps. Anisotropic pressure was applied using the Parrinello-Rahman barostat \cite{parrinello81}  with a time constant of 2 ps on the three different sides of the simulation box to allow independent fluctuations and changes in the shape of the solid region, avoiding possible stress in the solid. Cutoff radii of 9 \r{A} were used for van der Waals and electrostatic interactions. Long-range energy and pressure corrections were also applied to the Lennard-Jones part of the potential. The smooth Particle Mesh Ewald (PME) method \cite{essmann95} was employed to account for long-range electrostatic forces. The geometry of water molecules was constrained using the LINCS algorithm \cite{hess97,hess08b}. We used the well-known TIP4P/Ice \cite{JCP_2005_122_234511} and TIP4P/2005 \cite{abascal05b} potentials to describe water molecule interactions in our systems. Methane molecules were defined as a single Lennard-Jones site using parameters proposed by Guillot and Guissuni \cite{JCP_1993_99_8075} and Paschek \cite{JCP_2004_120_6674} (i.e., $\sigma$=0.373 nm and $\epsilon$=1.2264 kJ/mol). For the cross-interaction between TIP4P/Ice water and methane models, Lorentz-Berthelot rules were applied. When using the TIP4P/2005 model, we applied a positive deviation of 7$\%$ to the Lorentz-Berthelot energetic rule between water and methane.

\begin{table*}
\caption{\label{tabla-moleculas}\justifying{
Initial number of molecules for each phase (hydrate phase, liquid phase and gas phase) in the different configurations studied in this work. For the hydrate phase, the replication factor of the unit cell is indicated in each case. Excess methane is defined as the difference between the number of methane molecules in the gas phase and the number of methane molecules required for the complete growth of the hydrate, considering the water molecules in the liquid phase. For stoichiometric systems, the excess methane is zero. The next column indicates whether the system composition is stoichiometric. The last column stands for the system size in nm.}}
\begin{tabular}{c c c c c c c c c c c c c c c c c c c c c c}
\hline
\hline
Configuration & &   \multicolumn{5}{c}{Hydrate phase} & &
{Liquid phase} & &
\multicolumn{3}{c}{Gas phase} & & Stoichiometric & & System Size \\
     \cline{3-7} \cline{11-13}
     & \, & Unit Cell & \, & Water & \, & Methane & \, & Water &  \,  & Methane  &  \,  & Excess Methane & \, &  &\,& nm\\
\hline
1 & & 2$\times$2$\times$2 && 368 & &  64  && 368  & & 64  & & 0 & & Yes & & 5.5$\times$2.4$\times$2.4 \\
2 & & 2$\times$2$\times$2 && 368 & &  64  & & 368  & & 128 & & 64 & & No & & 6.9$\times$2.4$\times$2.4 \\
3 & & 2$\times$2$\times$2 && 368 & &  64  & & 368  & & 256 & & 192 & & No & & 9.6$\times$2.4$\times$2.4 \\
4 & & 2$\times$2$\times$2 && 368 & &  64  & & 736  & & 256 & & 128 & & No & & 12.2$\times$2.4$\times$2.4 \\
5 & & 4$\times$2$\times$2 && 736 & &  128 & & 736  & & 128 & & 0 & & Yes & & 10.9$\times$2.4$\times$2.4 \\
6 & & 3$\times$3$\times$3 & & 1242 & &  216  & & 1242  & & 216 & & 0 & & Yes & & 7.9$\times$3.6$\times$3.6 \\
7 & & 3$\times$3$\times$3 & & 1242 & &  216  & & 1242  & & 400 & & 184 & & No & & 12.0$\times$3.6$\times$3.6 \\
8 & & 3$\times$3$\times$3 & & 1242 & &  216  & & 2484  & & 432 & & 0 & & Yes & & 12.3$\times$3.6$\times$3.6 \\
9 & & 5$\times$5$\times$5 & & 5750 & &  1000 & & 5750  & & 1000 & & 0 & & Yes & & 13.1$\times$5.9$\times$5.9 \\
\hline
\hline
\end{tabular}
\end{table*}


\section{Results}

As previously mentioned, the first estimation of the three-phase coexistence temperature ($T_3$) for methane hydrate through computer simulation was performed in 2010 by some of us \cite{JCP_2010_133_064507}. We estimated the $T_3$ for a system of methane hydrate, liquid water, and methane gas in coexistence, with a central phase surrounded on each side by the other two phases. This arrangement facilitated three-phase coexistence due to periodic boundary conditions. This setup is analogous to those proposed in this work. $T_3$ was calculated for the TIP4P/Ice, TIP4P/2005, and TIP4P potential models. Maintaining the size of the solid methane hydrate slab, we studied both small and large systems by varying the number of molecules in the methane gas phase. In addition to the pioneering work of Conde and Vega, subsequent studies have provided various estimates of $T_3$. A compilation of these values is presented in Table \ref{tabla-pure}. Significant differences exist among reported $T_3$ values for methane hydrate, with key hypotheses attributing these differences to factors such as system sizes, initial configuration setups, or cutoff radius. The present work focuses on exhaustively studying the finite-size effects in the determination of $T_3$, maintaining the initial three-phase coexistence setup and fixing the cutoff radius value in all studied systems. We explore nine different configurations, varying the number of molecules in each of the three phases, aiming for a comprehensive understanding of the role of finite-size effects.

\begin{figure}[htp]
    \centering
     \includegraphics[width=0.5\textwidth]{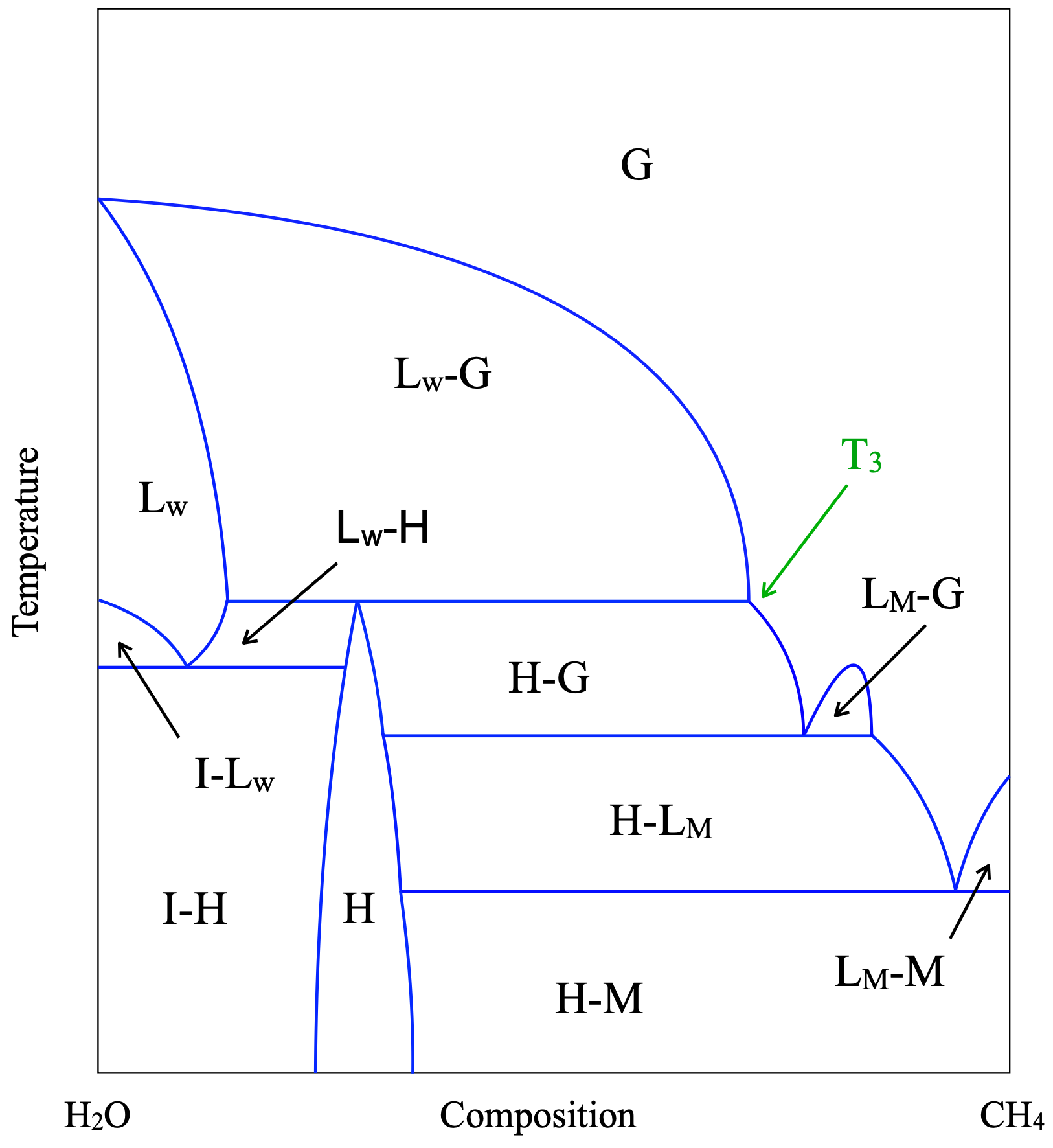}
    \caption{
    Artistical reproduction of the experimental CH$_4$-H$_2$O $T-x$ phase diagram at approximately 400 bar, illustrating the phases: hydrate (H), ice (I), liquid water (L\textsubscript{W}), liquid methane (L\textsubscript{M}), solid methane (M), and methane gas (G). The $T_3$ line represents the three-phase coexistence temperature (H-L\textsubscript{W}-V). 
    }
    \label{diagrama}
\end{figure}  

In accordance with the $T-x$ phase diagram for the CH$_4$-H$_2$O system (see Fig.~\ref{diagrama}), at temperatures below $T_3$, the system can evolve into different scenarios depending on the number of molecules in the initial configuration. 
A subtle issue should be mentioned, when the coexisting phases are changed (e.g., from ice to liquid water, the coexistence curve is continuous but the slope is not).
If the number of molecules in the liquid phase and the methane gas phase corresponds to the growth of hydrate in stoichiometric quantities (i.e., all hydrate cages occupied by a methane molecule), the system will evolve into a single phase of methane hydrate. If the reservoir of methane molecules in the gas phase exceeds the number of molecules in the liquid phase required for stoichiometric hydrate growth, the system will evolve below $T_3$ into two phases: methane hydrate in equilibrium with a methane gas phase. Conversely, if the reservoir of methane gas molecules is lower, there will be an excess of molecules in the liquid water phase once all possible methane hydrate has grown, leading the system to evolve at temperatures below $T_3$ into two phases: methane hydrate and liquid water. 

In this study, we explore two possible scenarios at temperatures below $T_3$: starting from stoichiometric configurations that evolve into a single phase of methane hydrate (configurations 1, 5, 6, 8 and 9), and from configurations with a higher number of methane molecules in the gas phase reservoir that will evolve into two phases, methane hydrate and methane gas (configurations 2, 3, 4, and 7).


\begin{figure*}
     \centering
     \begin{subfigure}[hbt]{0.30\textwidth}
         \centering
         \includegraphics[width=\textwidth]{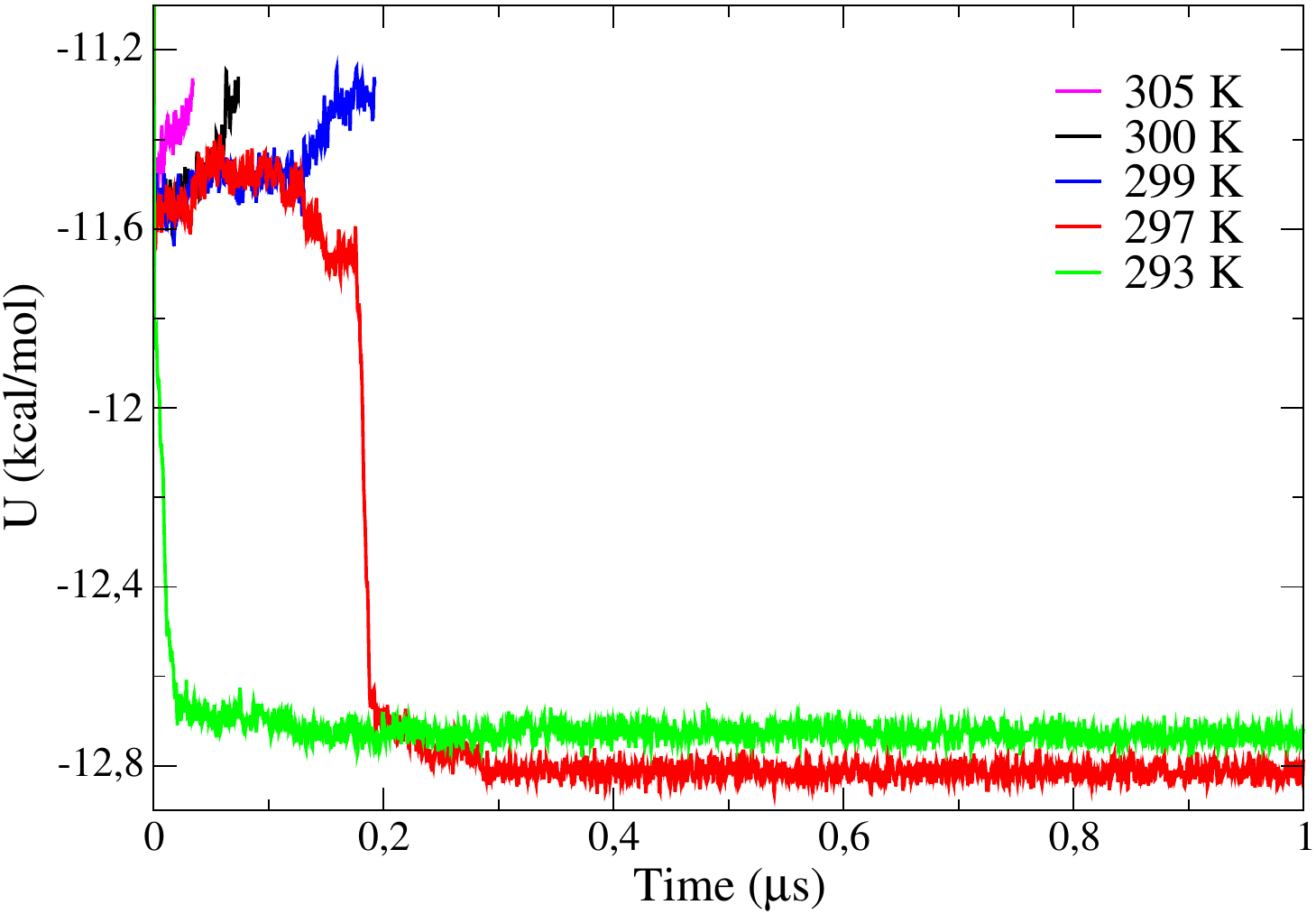}
         \caption{Conf. 1}
         \label{conf1}
     \end{subfigure}
     \hfill
     \begin{subfigure}[hbt]{0.30\textwidth}
         \centering
         \includegraphics[width=\textwidth]{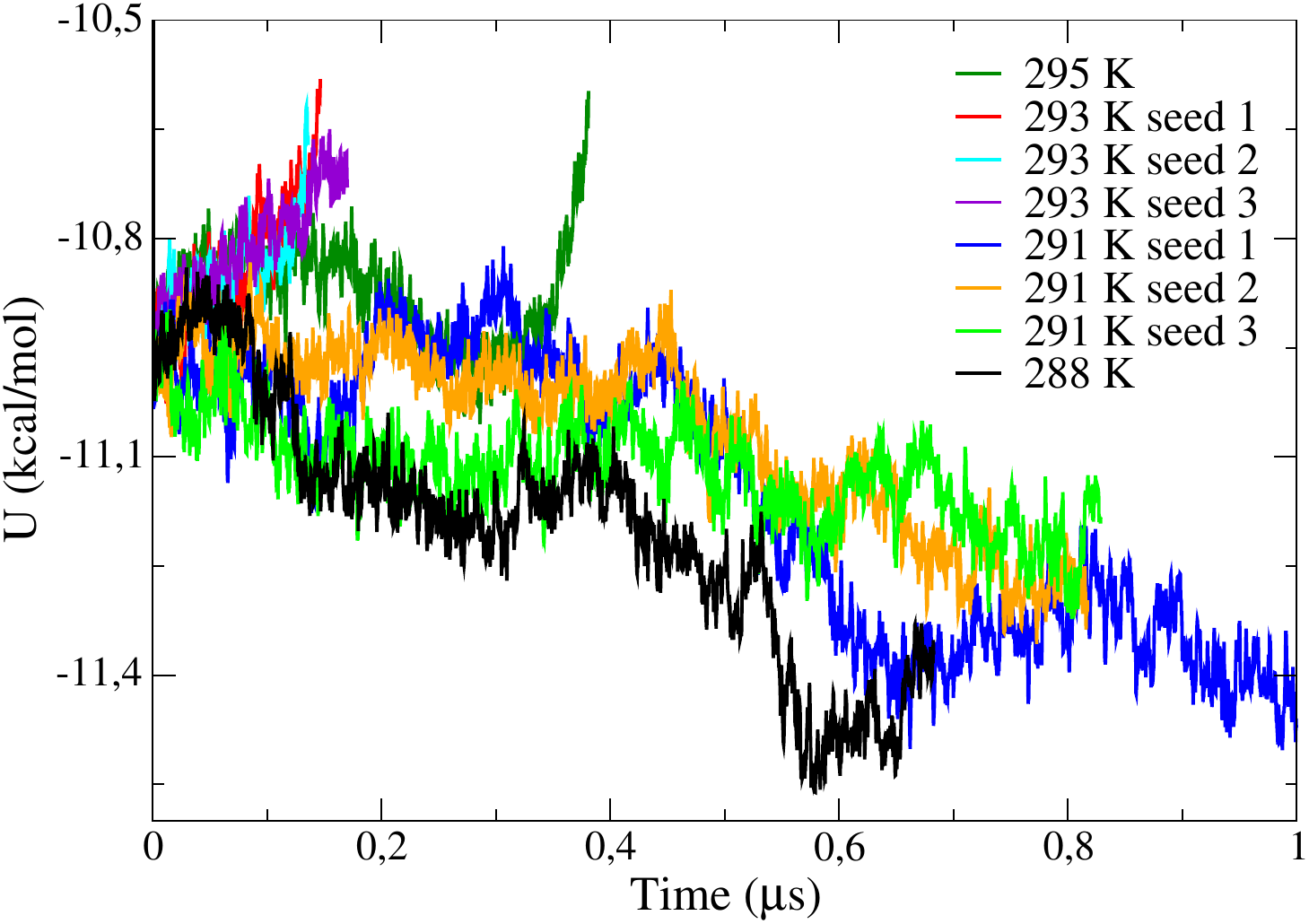}
         \caption{Conf. 2}
         \label{conf2}
     \end{subfigure}
     \hfill
     \begin{subfigure}[hbt]{0.30\textwidth}
         \centering
         \includegraphics[width=\textwidth]{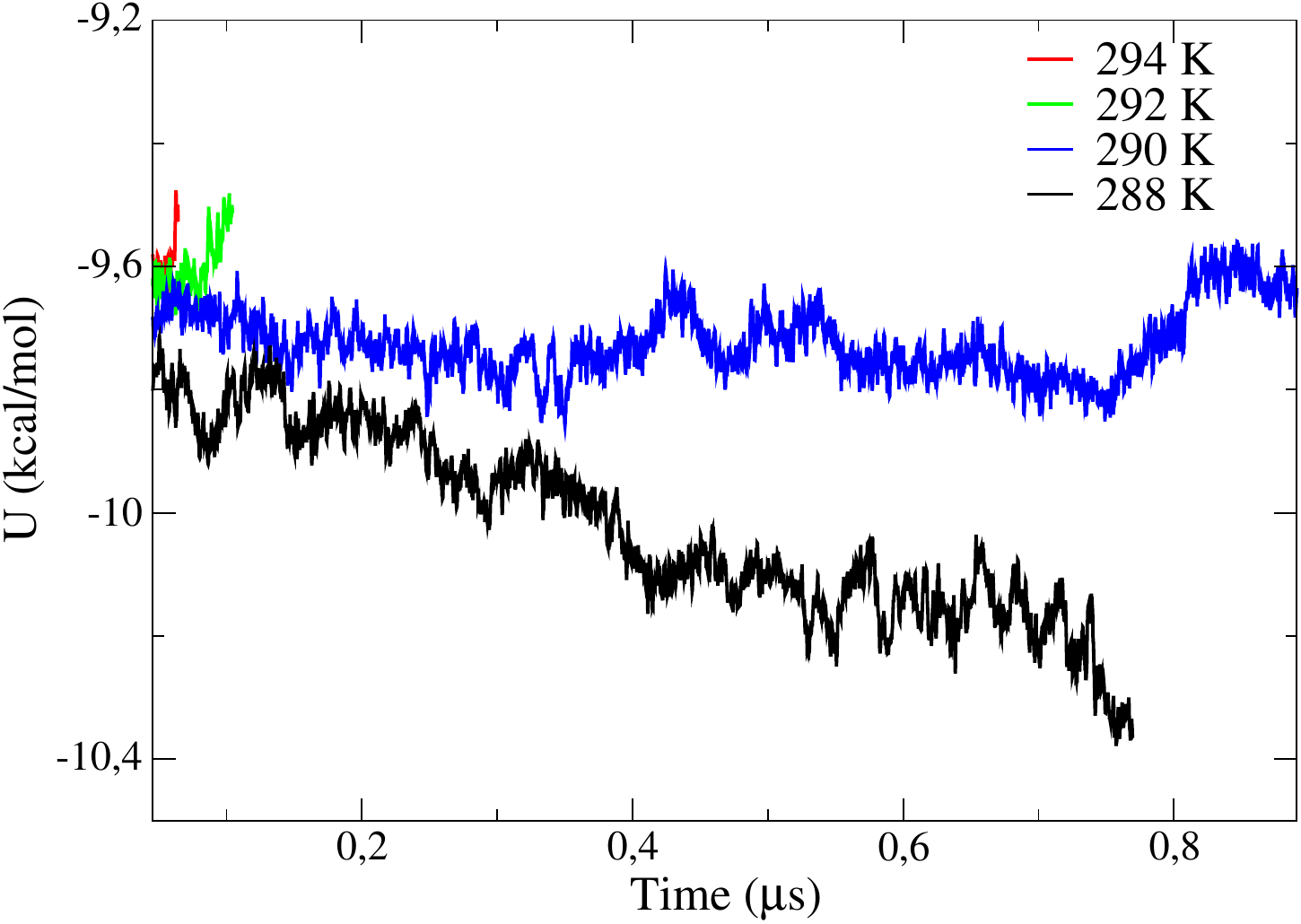}
         \caption{Conf. 3}
         \label{conf3}
     \end{subfigure}
     \hfill
     \begin{subfigure}[hbt]{0.30\textwidth}
         \centering
         \includegraphics[width=\textwidth]{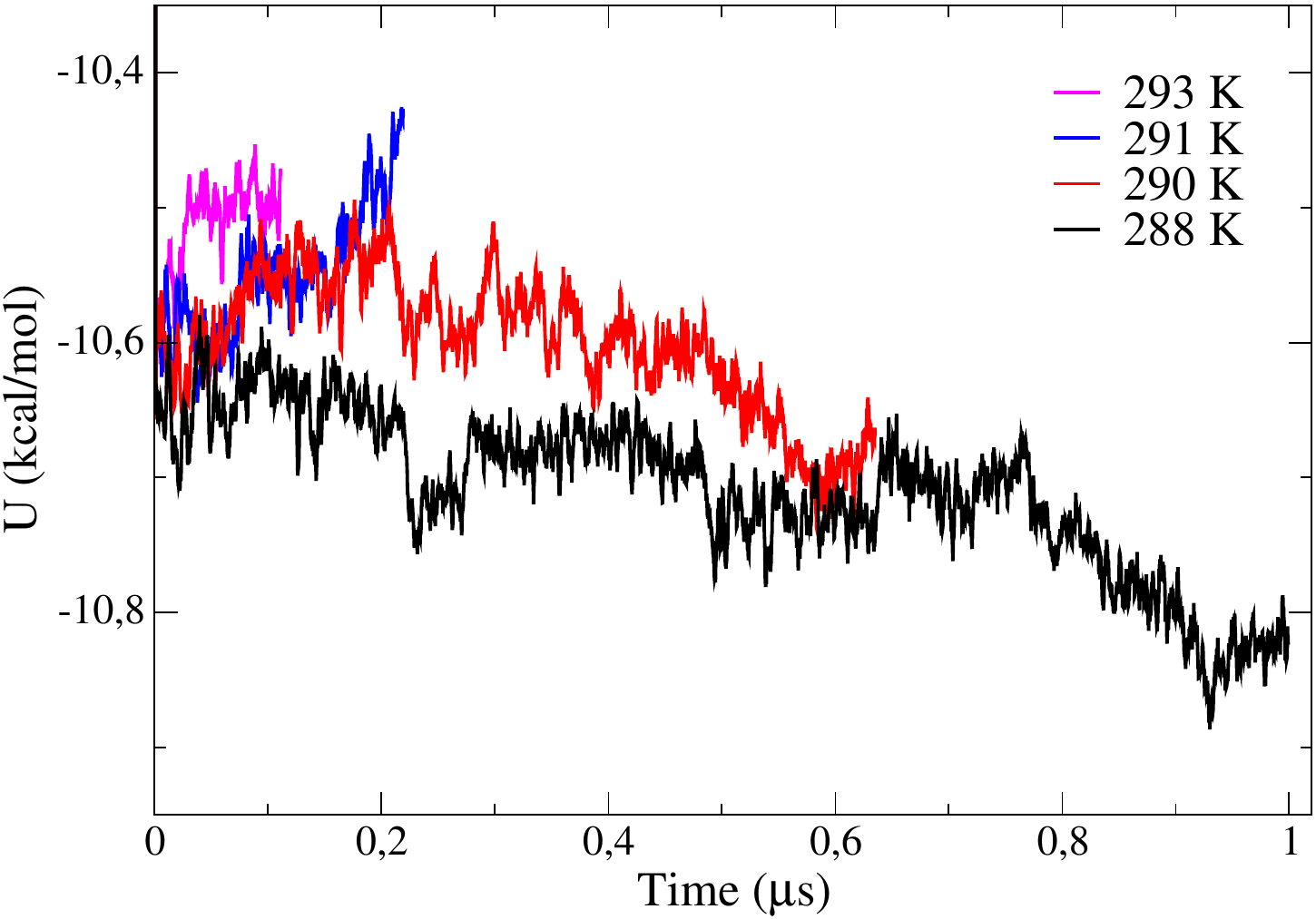}
         \caption{Conf. 4}
         \label{conf4}
     \end{subfigure}
     \hfill
     \begin{subfigure}[hbt]{0.30\textwidth}
         \centering
         \includegraphics[width=\textwidth]{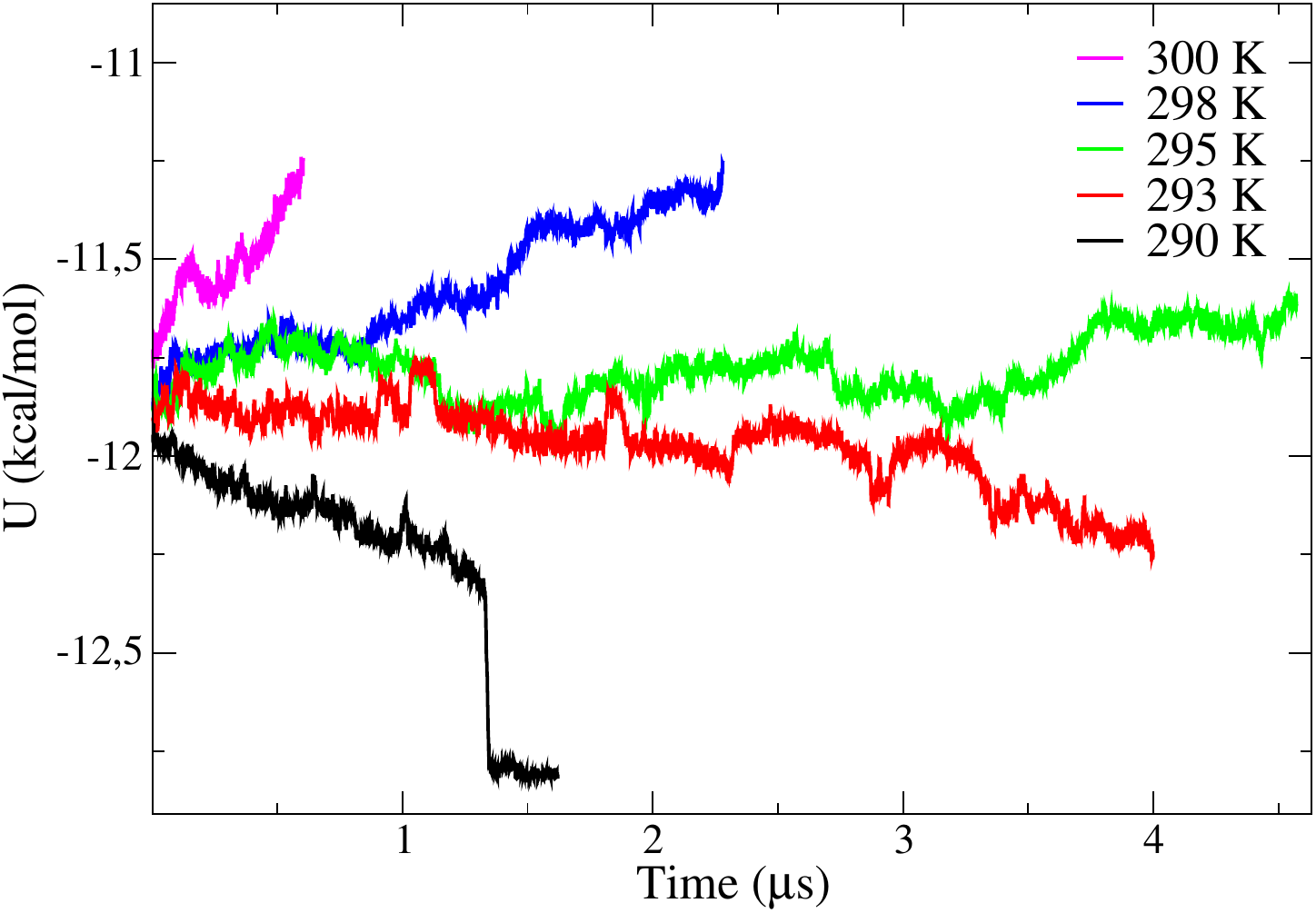}
         \caption{Conf. 5}
         \label{conf5}
     \end{subfigure}
     \hfill
     \begin{subfigure}[hbt]{0.30\textwidth}
         \centering
         \includegraphics[width=\textwidth]{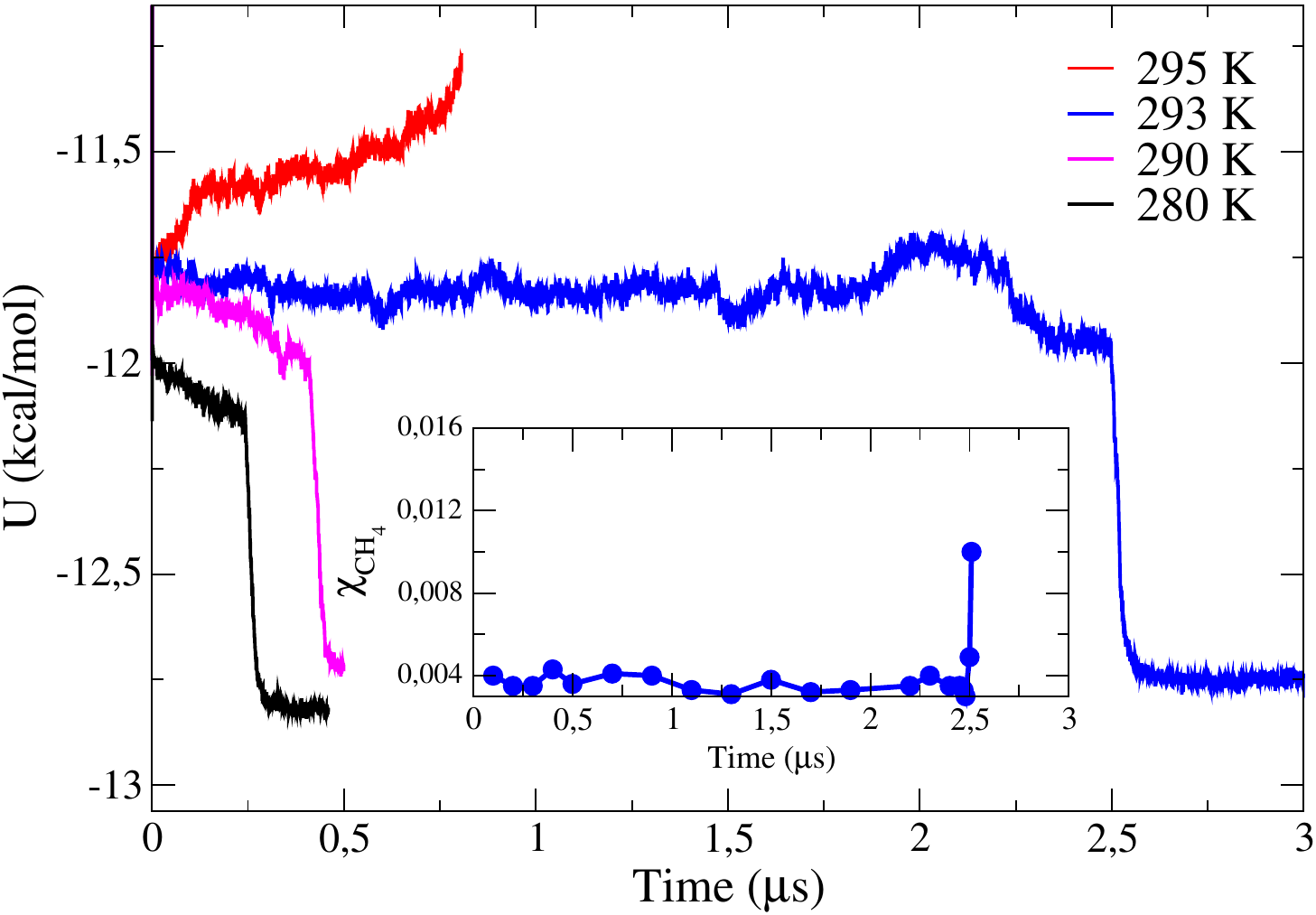}
         \caption{Conf. 6}
         \label{conf6}
     \end{subfigure}
     \hfill
      \begin{subfigure}[hbt]{0.30\textwidth}
         \centering
         \includegraphics[width=\textwidth]{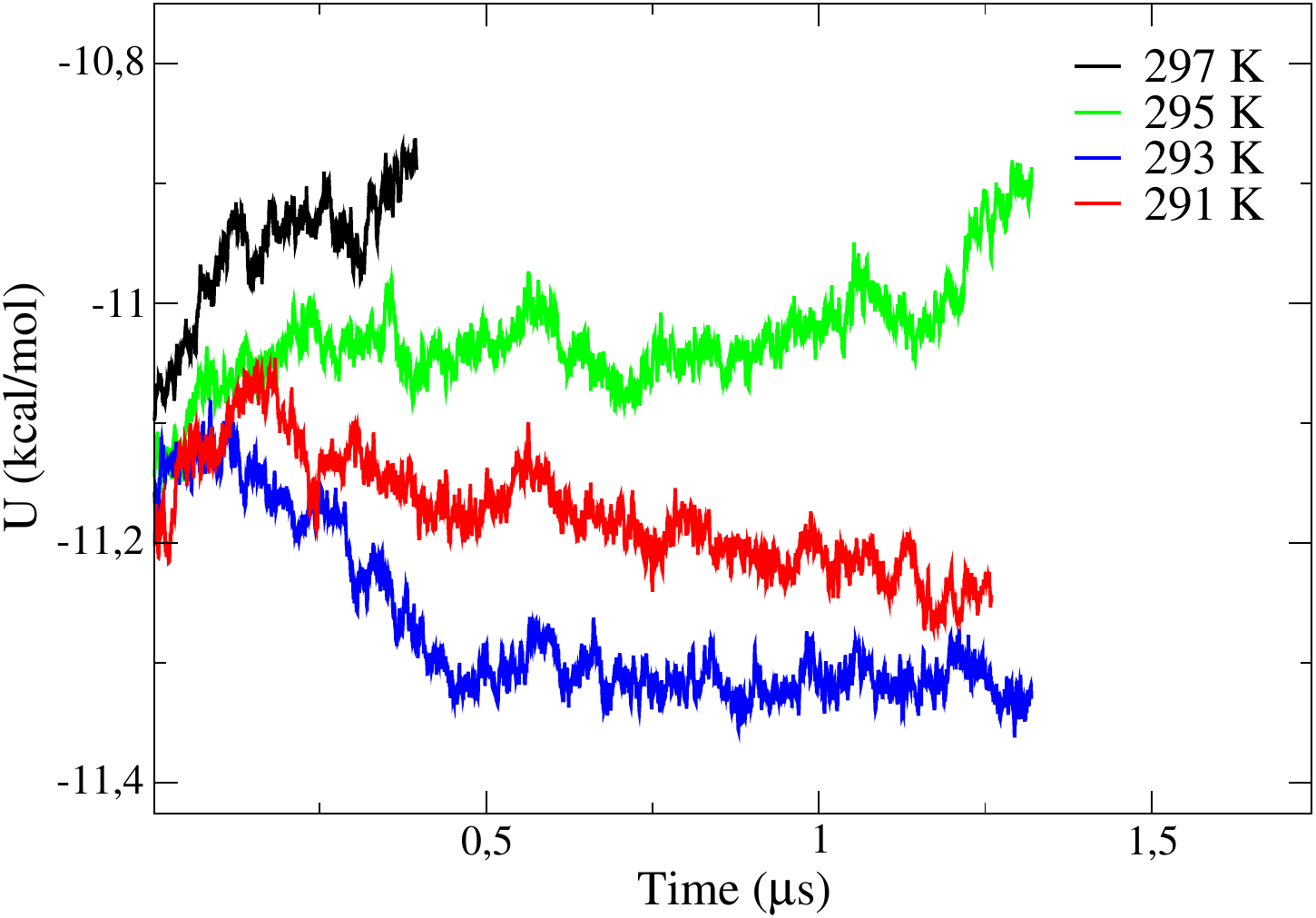}
         \caption{Conf. 7}
         \label{conf7}
     \end{subfigure}
     \hfill
     \begin{subfigure}[hbt]{0.30\textwidth}
         \centering
         \includegraphics[width=\textwidth]{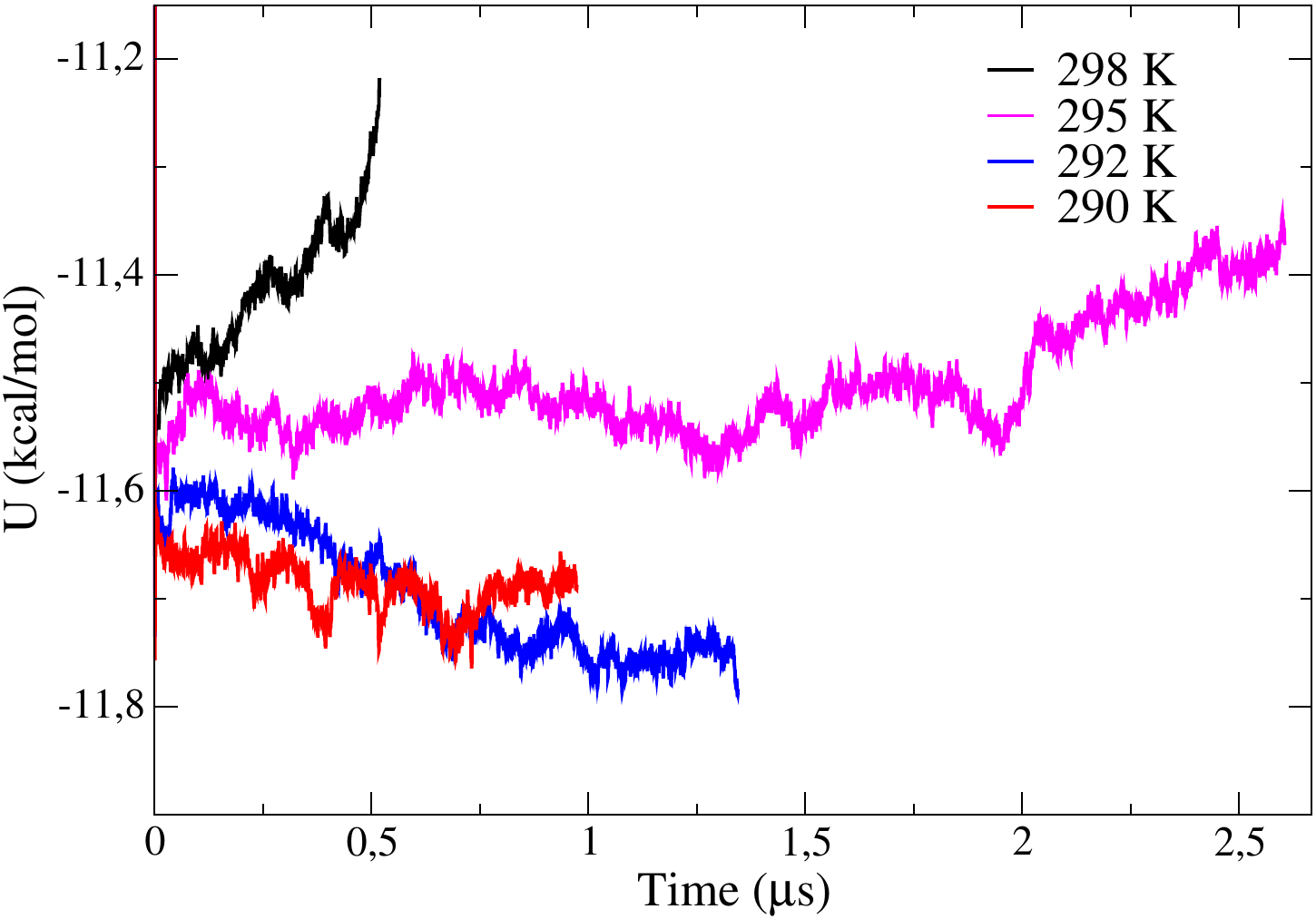}
         \caption{Conf. 8}
         \label{conf8}
     \end{subfigure}
     \hfill
      \begin{subfigure}[hbt]{0.30\textwidth}
         \centering
         \includegraphics[width=\textwidth]{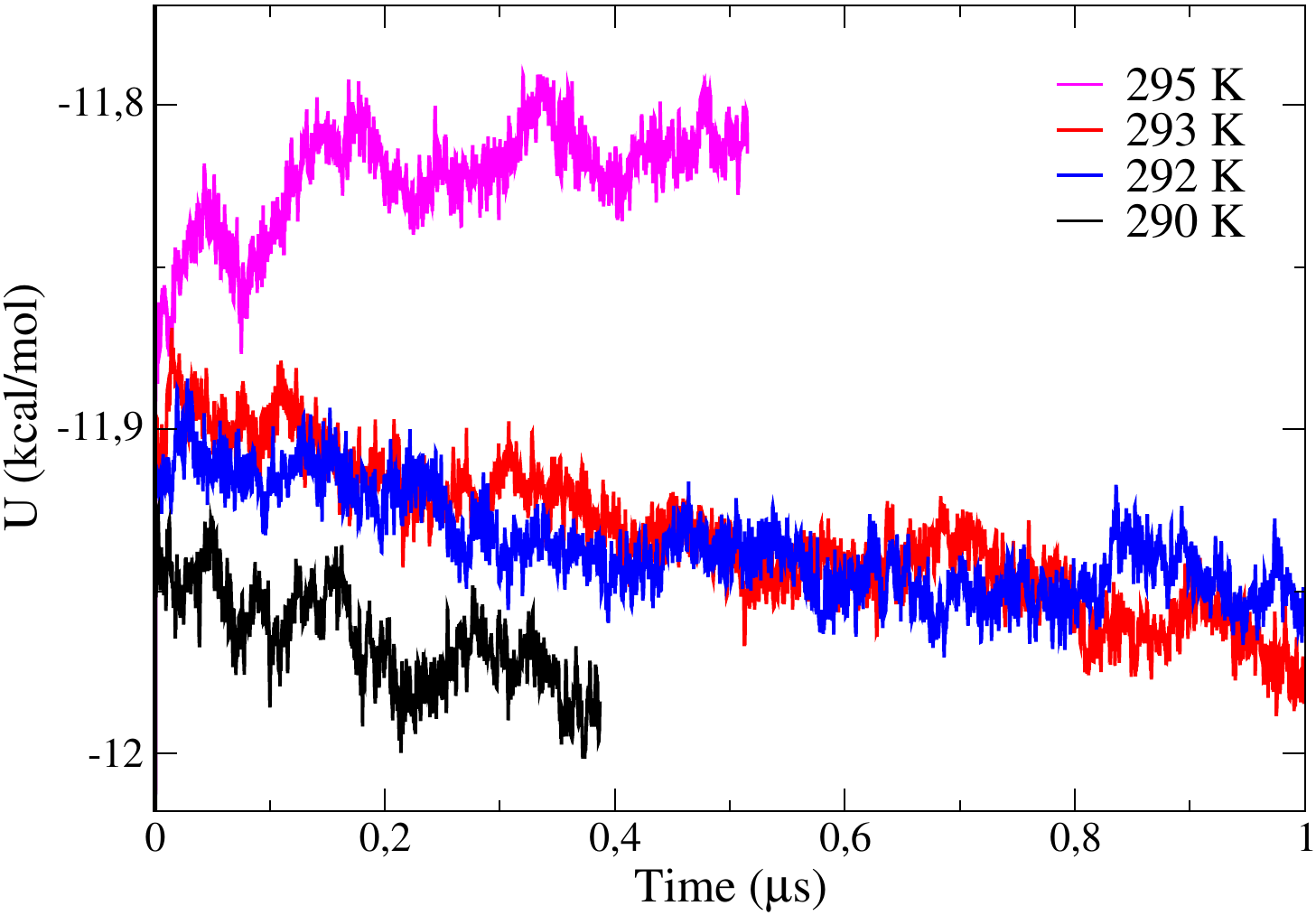}
         \caption{Conf. 9}
         \label{conf9}
     \end{subfigure}
        \caption{
\justifying{Evolution of the potential energy over time for the 9 size-dependent configurations analyzed in this work. The results were obtained from $NpT$ simulations at 400~bar using the TIP4P/Ice model. Notice that we have computed (see the inset of Fig. 2f) the concentration of the guest molecule (i.e., methane) in the aqueous phase as a function of time for 293 K.}}
        \label{confs-ice}
\end{figure*}

\begin{figure*}[htp]
    \centering
    \includegraphics[scale=0.4]{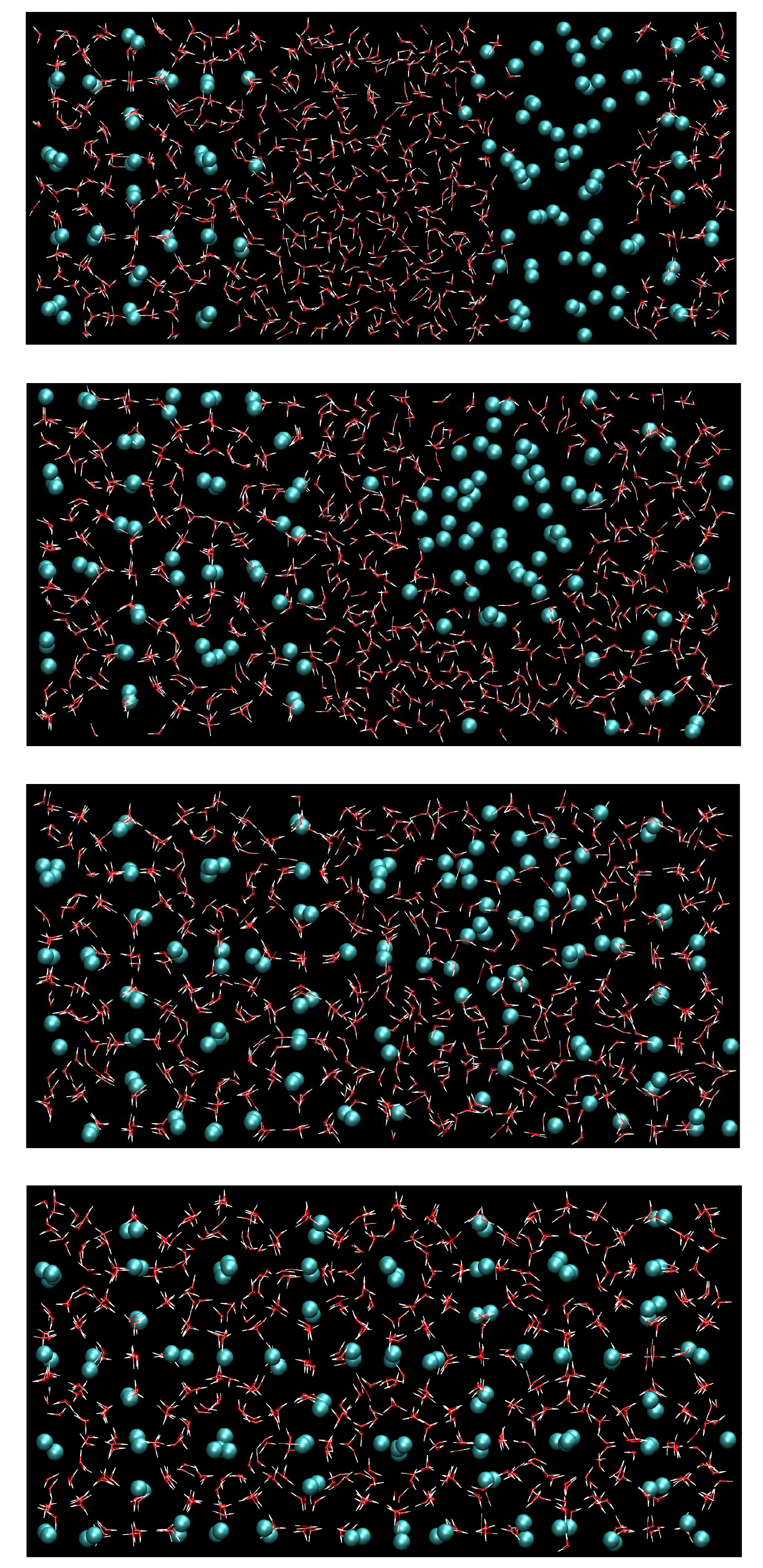}    
    \caption{\justifying{
    Snapshots taken at different times during a 1~$\mu$s simulation run for configuration 1 using the TIP4P/Ice model at 297~K and 400~bar. The formation of a bubble and the growth of methane hydrate are shown. From top to bottom: $t=0$~ns, three-phase initial configuration. $t=190$~ns, formation of a methane bubble within the liquid phase. $t=200$~ns, rupture of the methane bubble and the formation of an oversaturated solution. $t=270$~ns, complete growth of the methane hydrate. Water molecules are represented as white and red sticks and methane molecules as cyan spheres.
    }}
    \label{bubble1}
\end{figure*}


The first configuration we evaluate (configuration 1) is the same as the one studied by Conde and Vega \cite{JCP_2010_133_064507} in 2010, labeled as system A. As shown in Fig.~\ref{conf1}), at temperatures of 299, 300, and 305~K, the potential energy of the systems increases over time, indicating the melting of methane hydrate. On the contrary, at temperatures of 293 and 297~K, the potential energy of the systems undergoes a sharp decrease, indicating the growth of the hydrate. Considering these results, the three-phase coexistence temperature can be estimated between 299~K and 297~K. Thus, $T_3$ for configuration 1 is 298(1)~K. Conde and Vega reported a value of 302(3)~K for the same system size, which is consistent (within the error bar) with our result obtained 14 years later. It is worth noting that in this study, our simulations are on the order of hundreds of nanoseconds to microseconds, whereas in the previous study, we were on the order of tens of nanoseconds. This difference in simulation times could potentially explain the 3~K discrepancy in the $T_3$ value.

Certainly, the probability of selecting the exact value of $T_3$ in a simulation (with all its significant figures) is zero. Therefore, in practice, the system at a fixed pressure will increase or decrease its potential energy, estimating $T_3$ as the intermediate value between the lowest temperature at which the methane hydrate melts and the highest temperature at which the methane hydrate grows. As the system approaches the three-phase coexistence temperature, its behavior becomes increasingly stochastic and may require longer simulation times. For instance, in Fig.~\ref{conf1} at 297~K, the initial 100~ns of the simulation run show an increase in the potential energy, suggesting a trend towards hydrate melting. However, with extended simulation time, the potential energy subsequently decreases, signifying the complete growth of the hydrate phase. In complex systems like these, minor fluctuations can lead to either the melting or growth of the methane hydrate phase. It is imperative to allow sufficient time to ensure a thorough understanding of the system evolution. While achieving a more precise determination of $T_3$ would involve performing simulations for each temperature from different initial configuration seeds, similar to the methodology employed in ice systems \cite{conde2017high}. This approach exceeds the scope of our current work, which primarily aims to study the role of finite-size effects. In all cases, we provide a margin of error of at least 1~K.
In any case, notice that the error bars of the results have been estimated as the difference between the calculated $T_3$ and the closest above and below temperatures. It is true that if one wants to be rigorous several trajectories for the same temperature should be run  (as we will exemplify in Fig. \ref{conf2}) however,  the computational cost of running several trajectories for each temperature is too much demanding for the objectives of this work.


Given that configuration 1 corresponds to a stoichiometric configuration, the decrease in potential energy implies the evolution of the system into a singular phase of methane hydrate. In Fig.~\ref{conf1}, it is evident that at temperatures below $T_3$, there is a sharp decrease in potential energy until reaching a plateau where the potential energy stabilizes. This behavior not only signifies the complete growth of the methane hydrate phase (indicated by constant potential energy) but also unveils the formation of a methane bubble within the liquid phase (resulting in an abrupt decrease in potential energy).

This bubble formation was observed previously by other authors. In fact, Walsh \textit{et al.} \cite{doi:10.1021/jp206483q,Walsh1095} calculated nucleation rates after observing spontaneous nucleation of methane hydrate preceded by a bubble formation. In the determination of the $T_3$ performed by Conde and Vega in 2010 they also observed bubble formation before hydrate growth. In 2011, this bubble formation was also shown by Liang and Kusalik \cite{liang2011exploring} for H$_2$S systems. After these pioneering studies, other works have studied the effect of bubble formation in the  dissociation temperature of methane hydrates \cite{JPCB_118_1900_2014,bagherzadeh2015formation,grabowska2022solubility,fang2023effects}.
In fact, recently Grabowska \textit{et al.} \cite{grabowska2022solubility}, 
demonstrated that bubbles can be a way to obtain supersaturated solutions of methane and facilitate hydrate nucleation. In Fig.~\ref{bubble1}, we present a sequence of snapshots for configuration 1 at different times illustrating the growth of the hydrate phase at 297~K and 400~bar. As shown, starting from an initial three-phase configuration, we observe the gradual growth of a complete hydrate layer. After 190 ns, a methane bubble forms within the liquid phase, coinciding with the onset of the abrupt drop in the potential energy of the system at this temperature (see Fig.~\ref{conf1}). This bubble remains in motion for approximately 10~ns until it ruptures, generating supersaturation in the solution and causing fast hydrate growth. The rupture of the bubble coincides with the exact moment when the energy drop reaches its completion. Finally, the last snapshot shows the methane hydrate phase after its complete growth, where its potential energy stabilizes. The times at which each event occurs depend on the simulation temperature, increasing to larger system sizes and temperatures close to $T_3$. This observation suggests that bubble formation in this system significantly aids the fast growth of methane hydrate by promoting methane supersaturation in the liquid phase (limiting stage in hydrate growth). As discussed later, the formation of this bubble in small-sized systems with stoichiometric composition can lead to an erroneous determination of $T_3$ due to increased methane solubility in water. The value of $T_3$ and whether there is hydrate phase growth due to bubble formation are presented in Table \ref{tabla-t3-todos}.


The subsequent configuration we have studied consists of the same number of molecules in the hydrate and liquid phases, while the number of methane molecules in the gas phase differs (specifically, it is twice the number of molecules in configuration 1, as shown in Table \ref{tabla-moleculas}). For this new system, significant differences in the evolution of potential energy can be observed (see Fig.~\ref{conf2}). Firstly, the temperatures at which methane hydrate melts or grows markedly differ from those observed in the previous scenario, resulting in an estimated $T_3$ of 292(1)~K. 
In this case we have also simulated three different trajectories (using different seed numbers for generating randomly the initial velocities  from a Maxwell Boltzman distribution) for the temperatures closer to the $T_3$ (i.e., 291 and 293 K) and we observe the same behavior in all the independent trajectories.
This value is 6~K lower than the $T_3$ obtained in configuration 1. Another notable difference is observed; at temperatures below $T_3$, the energy decreases continuously and gradually during the simulation run. Note that the time required for the hydrate to grow in configuration 2 is greater than in configuration 1. Additionally, there is no abrupt drop characteristic of bubble formation in configuration 2. The observation of the trajectories of our configurations confirms a gradual layer-by-layer growth of the hydrate phase without bubble formation. It is clear that the bubble appears when the number of methane molecules in the gas phase is stoichiometric with the number of water molecules in the liquid phase (i.e., if there are 368 water molecules in the liquid phase, 64 methane molecules are needed to complete the total growth of the hydrate phase).

Considering these differences, a question arises: Does increasing the number of methane molecules in the gas phase, in addition to preventing bubble formation, cause a progressively greater reduction in $T_3$? To answer this, we designed configuration 3 exactly like configuration 2 but with twice the number of molecules in the gas phase. The only difference between configurations 1, 2, and 3 is the size of the gas phase (64, 128, and 256, respectively). The energies of this configuration 3 are shown in Fig.~\ref{conf3}, and the $T_3$ is 289(1)~K. Similarly to configuration 2, in this case, the slow decrease in energy suggests no bubble formation in hydrate growth. However, although the $T_3$ from configuration 2 to configuration 3 undergoes a slight decrease, it is not as pronounced as the 6 K decrease from configuration 1 to configuration 2, indicating that bubble formation increases hydrate stability, resulting in a higher $T_3$.



We shall now examine the effect of increasing the number of molecules in the liquid phase. For that purpose, we have employed a new configuration (configuration 4) in which we keep the number of methane molecules in the gas phase and the size of the hydrate relative to configuration 3, but we double the number of water molecules in the liquid phase. As in the two previous cases, the number of methane molecules in the gas phase is not stoichiometric with the number of molecules of liquid water, thus avoiding bubble formation. In Fig.~\ref{conf4}, the graph illustrates the potential energy evolution over time for simulated temperatures in this configuration, yielding an estimated $T_3$ of 290(1) K. As expected, given its non-stoichiometric composition, the potential energy changes continuously and gradually during both hydrate melting and growth. Although with slight differences for the three non-stoichiometric configurations studied so far (configurations 2, 3 and 4), the value of $T_3$ is around 291 K, deviating significantly from the $T_3$ estimate for configuration 1 and the original work of Conde and Vega \cite{JCP_2010_133_064507}, where bubble presence was observed in the growth of the hydrate phase. From these findings, it can be concluded that increasing the number of molecules in both the liquid and gas phases, while maintaining a non-stoichiometric composition, prevents bubble formation and provides a consistently estimated $T_3$ value. However, further exploration is required to evaluate how the number of molecules in the hydrate phase may impact the $T_3$ value due to finite-size effects. Besides, not only the number of molecules in each phase, but also the cross-sectional area can be responsible of the discrepancies of $T_3$ for different systems. In fact, Rozmanov and Kusalik, showed the effect of cross-sectional area in the melting point of ice I$_h$ \cite{rozmanov2011temperature}.



We now proceed to increase the number of molecules in the solid phase of methane hydrate, which is the only phase that has remained constant in our study up to this point. Our initial approach involves doubling the number of molecules in all phases. To achieve this, we replicate the unit cell of configuration 1 by two along the $x$-axis, resulting in a 4$\times$2$\times$2 unit cell for configuration 5. Considering the composition of the phases, this configuration has a stoichiometric composition. It is essential to note that doubling the number of molecules in each phase significantly increases the computational cost, transitioning from simulations at the nanosecond scale to the microsecond scale. The results obtained in the simulation of this configuration are presented in Fig.~\ref{conf5}, revealing a significant increase in the simulation time required to observe the growth or melting of the hydrate. The three-phase equilibrium temperature for this system size ranges between 295 K (the lowest temperature showing an increase in potential energy) and 293 K (the highest temperature showing a decrease in potential energy), given a $T_3$ value of 294(1) K.

Once again, due to its stoichiometric composition, this configuration manifests the formation of a methane bubble, evident in the abrupt drop in potential energy at 290 K (see Fig.~\ref{conf5}). Bubble formation is also anticipated at 293 K, but its observation demands a more extended simulation time. These findings confirm that bubble formation during methane hydrate growth is caused by the stoichiometric composition of the system, regardless of its size. However, in this new configuration, the increase in the $T_3$ value is not as pronounced, indicating that discrepancies in $T_3$ values are attributable not to bubble formation but to the size of the system. To validate this hypothesis, we proceed to further increase the system size. The subsequent step involves designing a 3$\times$3$\times$3 unit cell for the hydrate phase, maintaining an equivalent number of molecules for the liquid and gas phases, resulting in configuration 6.

\begin{figure*}[htp]
    \centering
    \includegraphics[scale=0.4]{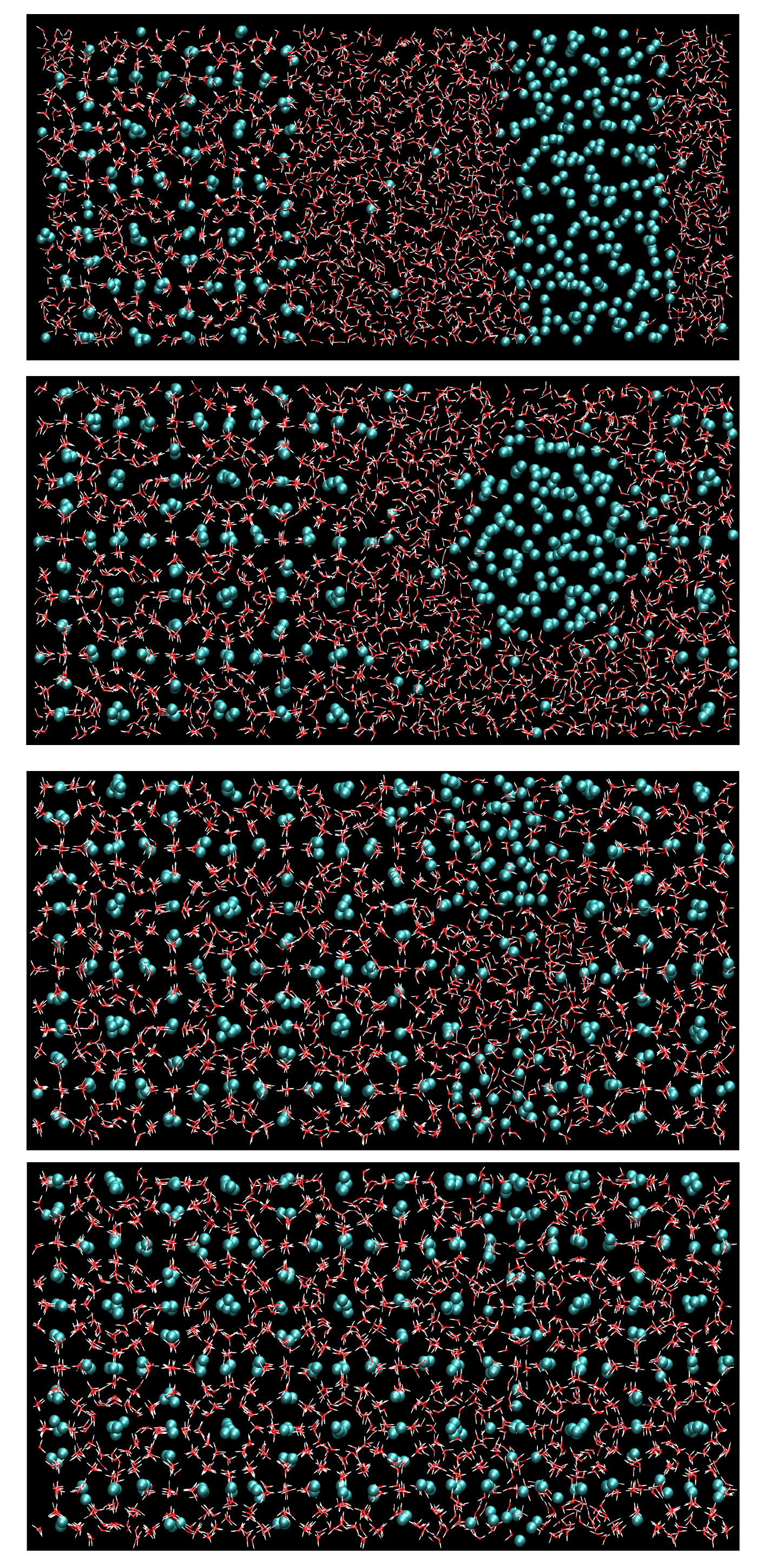}    
    \caption{\justifying{
    Snapshots taken at different times during a 500~ns simulation run for configuration 6 using the TIP4P/Ice model at 290~K and 400~bar. The formation of a bubble and the growth of methane hydrate are shown. From top to bottom: $t=50$~ns, some methane molecules from the gas phase incorporate into the liquid water phase. $t=415$~ns, formation of a methane bubble within the liquid phase. $t=440$~ns, rupture of the methane bubble and the formation of an oversaturated solution. $t=500$~ns, the methane hydrate is completely grown. Water molecules are represented as white and red sticks and methane molecules as cyan spheres.
    }}
    \label{bubble2}
\end{figure*}

As shown in Fig.~\ref{conf6}, configuration 6 displays a pronounced decrease in potential energy with a high slope at temperatures below $T_3$, indicating the formation of a methane bubble within the liquid phase. This event accelerates the complete growth of the methane hydrate phase. The abrupt decrease in potential energy is attributed to a significant increase in methane solubility in water, leading to an acceleration in the growth dynamics of methane hydrate. Conversely, configuration 7, sharing an identical 3$\times$3$\times$3 unit cell with configuration 6, differs in its global composition (see Table \ref{tabla-moleculas}). The increased number of methane molecules in the gas phase results in a non-stoichiometric configuration. Notably, configuration 7 avoids bubble formation and prevents overestimation of methane solubility in the liquid phase. For this configuration, the decrease in potential energy below $T_3$ is gradual (see Fig.~\ref{conf7}). The $T_3$ obtained for this size is the same as for configuration 6, with an estimated value of 294(1)~K. 
Likewise, in the case of the bubble formation (i.e., configuration 6, 293 K) we have computed the solubility of methane in the aqueous phase as a function of time, observing that just at the beginning of the bubble formation there is a large increment of methane concentration in the liquid (see inset of Fig. \ref{conf6}). This was also observed by Kusalik and coworkers for other hydrates\cite{hall2016unraveling}.
The complete growth sequence for the stoichiometric composition of configuration 6 is given in Fig.~\ref{bubble2}. The simulation run at 290~K, starting from an initial three-phase configuration, evolves to form a methane bubble within the liquid phase. The bubble formation occurs at the exact moment when the potential energy starts to drop abruptly (see Fig.~\ref{conf6}). Subsequently, the bubble decreases in size and the hydrate grows (coinciding with the drop in energy observed in Fig.~\ref{conf6}) until finally ruptures, leading to the appearance of a supersaturated solution and resulting in the final growth of the hydrate phase. This methane hydrate phase is the only stable phase at temperatures below $T_3$ due to its stoichiometric condition. This sequence of snapshots is analogous to that observed for configuration 1, with the difference that at larger sizes, the value of $T_3$ apparently is not influenced by the formation of the bubble since the system, before the appearance of the bubble, has undergone a gradual decrease in energy and consequently a partial growth of methane hydrate. In both configurations, the time of bubble formation and rupture coincides with the high slope of potential energy in Figs.~\ref{conf1} and \ref{conf6}. The final growth of the hydrate phase occurs at the end of the slope. Also, note that during this sharp energy decrease, the size of the bubble is decreasing, thereby increasing the solubility of methane. 

In the supplementary material, we provide a movie of the simulation trajectory at 290~K and 400~bar for configuration 6 using the TIP4P/Ice model. The movie illustrates the diffusion of methane molecules from the gas phase to the liquid phase and the formation of the bubble. It shows how the bubble gradually reduces in size (thus, growing the hydrate at the same time) until it ruptures, resulting in a supersaturated solution and the complete growth of the hydrate phase. The visualization of the trajectory reveals a curved interface between the methane bubble and liquid water, contrasting with the planar interfaces observed in the rest of the phase coexistence in the system.

\begin{figure}[htp]
    \centering
    \includegraphics[scale=0.35]{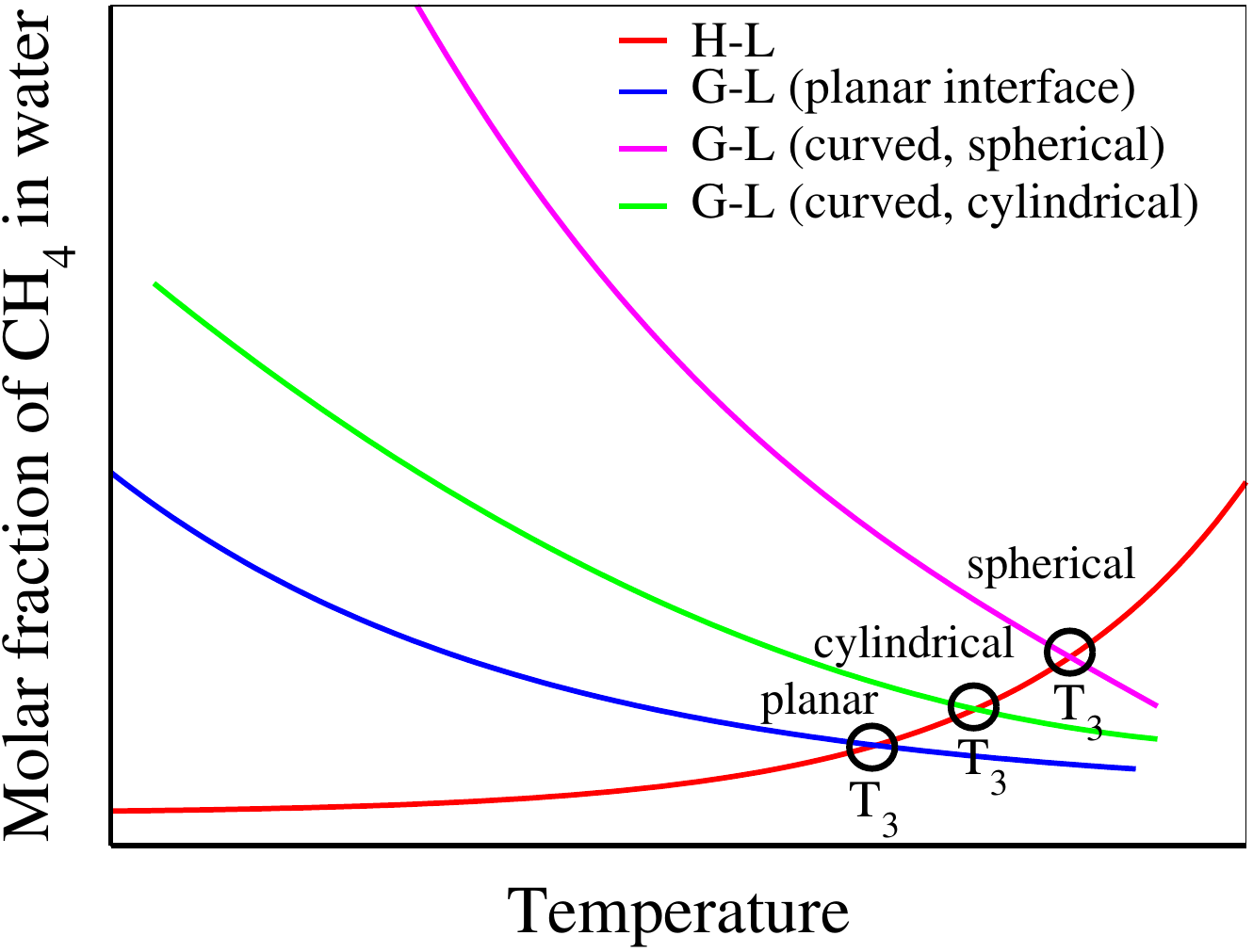}    
    \caption{\justifying{
    Schematic representation of methane solubility in the aqueous phase when in equilibrium with the gas phase  (G-L) with three different geometries for the G-L interface or when in equilibriun wih the hydrate solid  (H-L) \textit{via} a planar interface. The intersection of the two curves defines the three-phase coexistence temperature ($T_3$). We present variations in the methane solubility curve, considering different scenarios for the G-L interface: with a planar interface and a curved interface when methane is embedded within a spherical or cylindrical bubble. The presence of a methane bubble shifts $T_3$ to higher temperature values. }}
    \label{sketch}
\end{figure}

\begin{figure}[htp]
    \centering
    \includegraphics[scale=0.35]{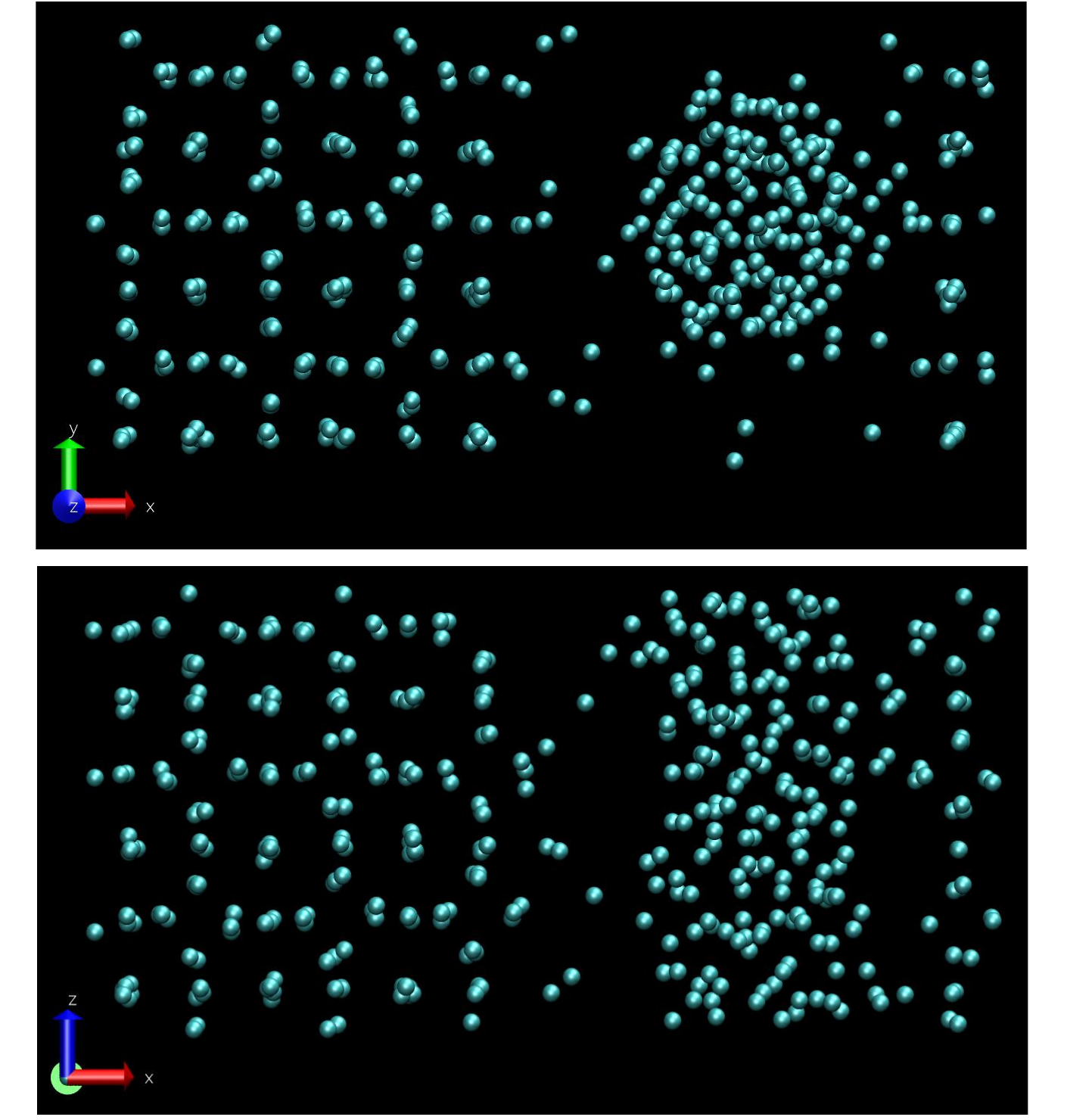}    
    \caption{\justifying{
    Snapshots of the $xy$ and $xz$ planes taken for configuration 6 using the TIP4P/Ice model at 290~K and 400~bar, showcasing the emergence of a methane bubble within the liquid phase at $t=450$~ns. For clarity, water molecules are omitted, while methane molecules are represented as cyan spheres. Methane molecules are visible within the hydrates on the left and right sides of the figure. In the central region, methane molecules are shown forming a cylindrical bubble.  
    }}
    \label{bubble-planos}
\end{figure}  


Upon analyzing the data, it initially appears that, at a certain size where finite-size effects become negligible, the presence or absence of a bubble does not affect the estimate of $T_3$. However, this interpretation is incorrect. To clarify this matter, it is crucial to note that the bubbles in our study are small, with radii of approximately 0.4 nm and 0.7 nm for the bubbles showed in Figs. \ref{bubble1} and \ref{bubble2}, respectively. These small bubbles are stable only for a limited time (i.e., 10-25 ns) in our case and besides their size is shrinking while the hydrate is growing reaching a finite size at which they become mechanically unstable. The scenario may change if the bubbles are larger and stable for longer times. In our recent work \cite{grabowska2022solubility}, we demonstrated that the $T_3$ can be calculated using the solubility method. This method involves evaluating the solubility of methane in both a liquid-gas system and a hydrate-liquid system as a function of temperature. The intersection of these two solubility curves determines $T_3$. However, it is important to emphasize that the three-phase coexistence temperature can vary with the curvature of the interface. In the case of a planar interface, the obtained $T_3$ is lower compared to a curved interface. This difference is attributed to the Laplace equation, which indicates that the solubility of a gas in a gas-liquid system is higher with a curved interface. Consequently, $T_3$ changes to higher values with a curved interface, as illustrated in Fig.~\ref{sketch}. The enhanced solubility of methane in the presence of bubbles compared to a planar interface at a constant temperature is straightforwardly explained by observing that the pressure within the cylindrical bubbles exceeds 400 bar (the outside pressure of the system), as indicated by the Laplace equation. This higher internal pressure yields higher chemical potentials, leading to a higher molar fraction assuming ideal behavior for methane in water. As in our previous work\cite{grabowska2022solubility} we can estimate the internal pressure by using the Laplace equation for a cylinder ($\Delta P = \frac{\gamma}{R}$), assuming an interfacial tension ($\gamma$) of 40 mJ/m$^2$ (reported in our previous work at 290 K), and taking into account that the radius for the observed bubbles ($R$) in this work are 0.4 and 0.7 nm (for Configurations 1 and 6 respectively), the internal pressures  are about 1400 and 970 bar respectively. As expected, this pressure is much higher than the 400 bar of the global system, leading to higher solubilities of methane. In fact, in our previous work, we showed the modification of the chemical potential when using curved interfaces compared to the value at planar interfaces. We demonstrated that the chemical potential of methane increases by approximately 1.6 k$_B$T units when employing curved interfaces (spherical bubbles of about  in nm in that case). This increase induces a heightened driving force for nucleation, equivalent to the effect of approximately 20 K of additional supercooling. In this way, it can be explained why the nucleation of  hydrates is considerably easier in brute force simulations when bubbles are present\cite{doi:10.1021/jp206483q,Walsh1095}.

Another significant distinction in our case is that bubbles, which spontaneously emerge in stoichiometric systems when sufficient molecules of the gas phase are incorporated into the hydrate, assume a cylindrical shape rather than a spherical one (see Fig.~\ref{bubble-planos} for an example of the bubble formed in configuration 6 at 290~K from different perspectives, where the cylindrical shape of the bubble is clearly observed). According to the Laplace equation, cylindrical interfaces result in lower solubilities than spherical ones (although still higher than planar interfaces). Therefore, if the bubbles were stable and sufficiently larger, the $T_3$ would be shifted to higher temperatures compared to a planar interface but lower temperatures compared to spherical bubbles.

\begin{figure}[htp]
    \centering
    \includegraphics[scale=0.35]{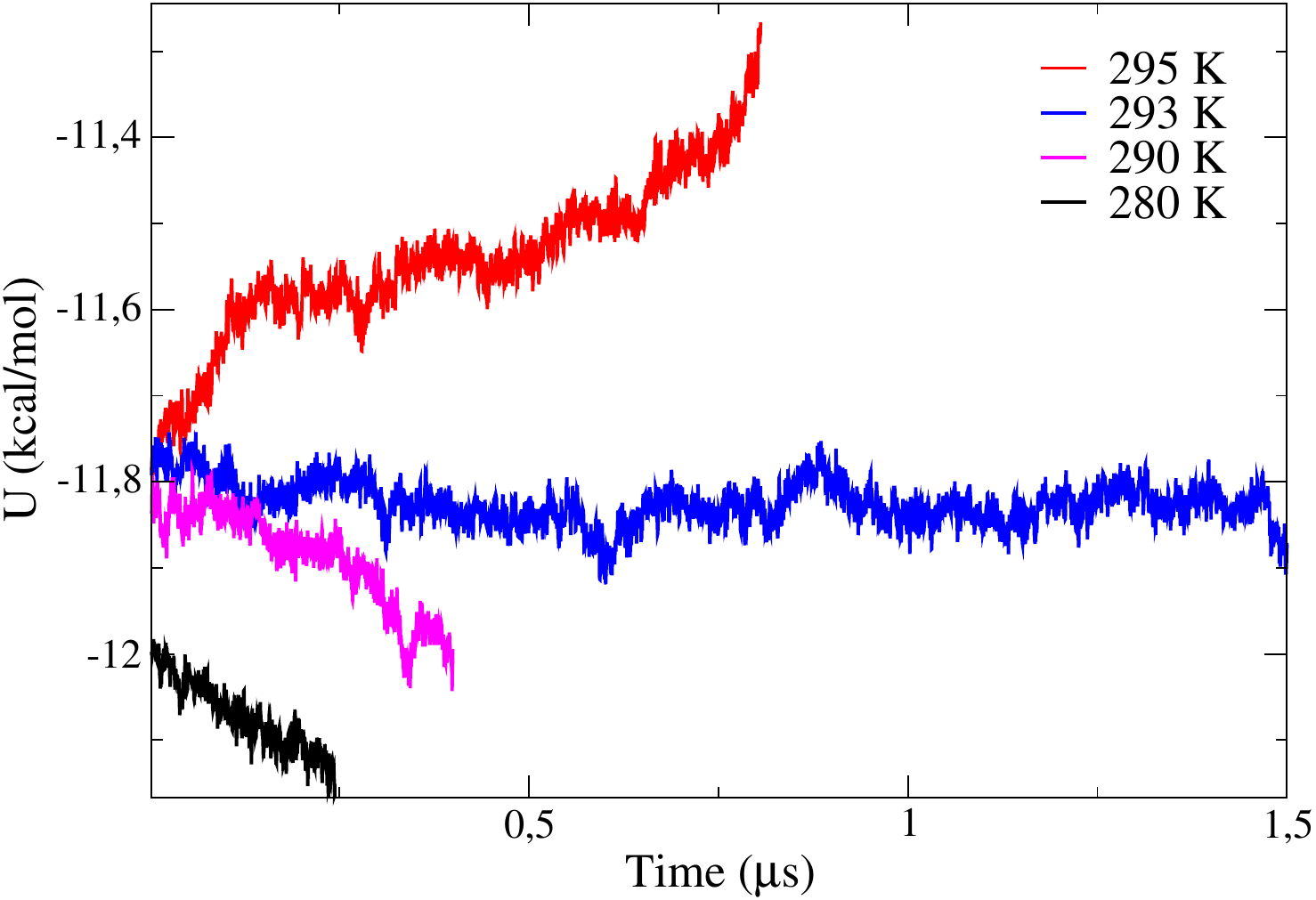}    
    \caption{\justifying{  
    Close-up view of Fig.~\ref{conf6} showing the temporal evolution of potential energy for configuration 6 analyzed in this study using the TIP4P/Ice model, spanning from the initiation of the simulation to the precise moment just before the onset of methane bubble formation.    
    }}
    \label{t3-ampliada}
\end{figure}

In this work, our small bubbles are stable for a limited time of about 10-25 ns, but if they were stable longer times, $T_3$ would shift to higher values. It is important to note that although the impact of the bubbles in the estimate of  $T_3$ may be not too large
considering them in the estimate of  $T_3$ may in general led to incorrect results. By examining Fig.~\ref{conf6}, one might wonder about the impact on energies if only values before the appearance of the bubbles are considered (i.e., before the energy drop). Conducting this exercise (see Fig.~\ref{t3-ampliada}) reveals a clear energy decrease at 280~K and 290~K, while at 293 K, there is a small decrease before the drop which is more difficult to identify if corresponds to a hydrate growth. However, the estimated value of $T_3$ from the information in Fig.~\ref{t3-ampliada} is in perfect agreement with the previously indicated value of 294(1)~K for this configuration 6. Moreover, it can be concluded that the presence of a bubble can alter the estimation of $T_3$ for these reasons. Therefore, we strongly recommend using a non-stoichiometric system to avoid these problems. Specifically, we advocate for the use of configuration 7, which involves an affordable number of molecules for simulation and avoids the formation of bubbles. Notice, that in case the rich-guest phase is more soluble, as in the case of CO$_2$, the number of molecules in this phase should be higher, for example using 800 molecules instead of 400.

Does this mean that stoichiometric systems would not be valid for determining $T_3$? No, they would be valid as long as the system is large enough (3$\times$3$\times$3 unit cells and beyond), and the trend of the potential energy curves clearly indicates, before the bubble formation (that is, before the abrupt decrease in potential energy), whether the system melts or grows. With this information and in retrospect, it is evident that the choice of configuration 1 used in our work in 2010 was not a good one due to the small size of the system (2$\times$2$\times$2 unit cell) and the rapid formation of the bubble, which did not allow us to determine whether the potential energy had increased or decreased before the bubble formation. Nevertheless, configuration 1 allowed us to obtain an initial estimate of $T_3$ with runs of about 200~ns and with a system of only 1000 molecules, which represented our limit of computational time and system size in 2010.

Finally, we explore the impact of finite-size effects by studying two larger configurations, 8 and 9, to determine their resilience to such effects. Configuration 8 maintained the $3\times3\times3$ unit cell but doubled the number of molecules in both the liquid and gas phases compared to configuration 6. On the other hand, configuration 9 featured a supercell $5\times5\times5$, resulting in a configuration with 13,500 molecules. Both configurations exhibited a stoichiometric composition. The estimated $T_3$ for both configurations was 294(2)~K. The uncertainty in these estimations increased compared to configurations with fewer molecules, due to the long simulation times required to observe the complete growth or fusion of the system. The slow dynamics only permitted the observation of slight fluctuations in potential energy (see Figs.~\ref{conf8} and \ref{conf9}). Nevertheless, these fluctuations allowed us to confidently estimate the $T_3$ for these large systems. To capture abrupt energy drops due to the stoichiometric condition of these systems, longer simulation times are required. However, the trends in the potential energy curves are sufficient to predict the behavior of the  system and estimate $T_3$. For both configurations, with stoichiometric composition, at temperatures below $T_3$ and with prolonged simulation times, the formation of a methane bubble in the liquid phase can be predicted. The bubble formation condition is always satisfied in stoichiometric systems, regardless of their size. For all runs with a stoichiometric composition of water and methane, the formation of bubbles is expected at the end of the run when few methane molecules remain in the gas phase, as long as the simulation is long enough.
Nevertheless, it is worth mentioning that the formation of the bubble is expected not only in systems satisfying  the stoichiometric composition  (i.e., when the ratio of molecules of methane in the gas phase to that of water in the liquid phase in the initial configuration is 8/46 , i.e., 0.174) but also in systems with a lower value of this ratio. In fact, in these cases, the formation of the bubble is expected to occur at shorter times. To further investigate this, we have simulated at 290 K direct coexistences of liquid water (1242 molecules) with a gas methane phase with different numbers of methane molecules (i.e., 50, 100, 150, 200, 300, and 400) during 50 ns. The simulations were performed applying a pressure of 400 bar in the Np$_x$T ensemble and the interfacial area of the systems was 12.96 nm$^2$. We observed the formation of the bubble only in the systems of 50 and 100 methane molecules. Thus, these findings reveal that, in this case, there is a critical thickness of the gas slab of about 0.8 nm (i.e., when the thickness of the gas phase is larger than 0.8 nm no bubble is formed in 50 ns). We have also checked the effect of the interface simulating during 50 ns a system with 5750 water and 1000 methane molecules with an interface of about 35 nm$^2$, but we do not observe bubble formation for this simulation time. In any case further work is needed to determine precisely under which conditions the planar water-methane interface is not stable with respect to the formation of a cylindrical or spherical bubble as was done for one component systems in other studies \cite{10.1063/1.2218845,10.1063/1.5097591,montero2022thermodynamics}. In fact, in the future, it would also be interesting to study the bubble shape in bigger systems.

Table~\ref{tabla-t3-todos} summarizes the findings for the 9 size-dependent configurations, presenting the $T_3$ values for each size and indicating whether a bubble forms. While it may seem that bubble formation consistently distorts the results, this is not universally true and only occurs in cases with smaller sizes. In fact, in larger systems, the formation of a bubble can accelerate the overall growth of the hydrate, leading to a fast descent in potential energy rather than a gradual one. Now, as mentioned previously, if one aims to accurately determine $T_3$ in stoichiometric systems, it is necessary to analyze the potential energy curves before the bubble formation occurs.

These findings confidently lead us to assert that, from configuration 6 with a $3\times3\times3$ unit cell onwards, the value of $T_3$ remains constant and is not influenced by finite-size effects, regardless of whether a bubble forms. Configuration 5 is also unaffected by finite-size effects, having increased the replication factor of the unit cell in one direction. Among all the configurations studied, the best choice in terms of size/computational time ratio would be the non-stoichiometric system of configuration 7. In smaller systems with a non-stoichiometric composition (configurations 2, 3, and 4), $T_3$ is affected by finite-size effects, leading to a slight underestimation of its value. However, unequivocally, the configuration that must be disregarded when accurately determining $T_3$ is configuration 1, with a $2\times2\times2$ unit cell and stoichiometric composition. In this case, the value of $T_3$ is erroneously overestimated due to the formation of a bubble in this small system size, falsely promoting hydrate growth and consequently yielding an artificially higher stability and $T_3$ value.

\begin{table}
\caption{\label{tabla-t3-todos}\justifying{
Three-phase coexistence temperatures ($T_3$) for methane hydrate systems at 400 bar for TIP4P/Ice and TIP4P/2005 models obtained for the 9 size-dependent configurations. The error bars are given in parentheses. The replication factor of the methane hydrate unit cell is shown to relate the configuration number to size. The last column indicates bubble formation during the methane hydrate growth process.
}}
\begin{tabular}{c c c c c c c }
\hline
\hline
Conf. & Unit Cell & & $T_3$ TIP4P/Ice & & $T_3$ TIP4P/2005 & Bubble \\
\hline
1 & 2$\times$2$\times$2 & & 298(1) & & 282(1) & Yes\\
2 & 2$\times$2$\times$2 & & 292(1) & & 277(1) & No\\
3 & 2$\times$2$\times$2 & & 289(1) & & 277(1) & No \\
4 & 2$\times$2$\times$2 & & 290(1) & & 277(1) & No\\
5 & 4$\times$2$\times$2 & & 294(1) & & 280.5(1) & Yes\\
6 & 3$\times$3$\times$3 & & 294(1) & & 280.5(1) & Yes\\
7 & 3$\times$3$\times$3 & & 294(1) & & 280(1) & No\\
8 & 3$\times$3$\times$3 & & 294(2) & & 280(1) & expected\\
9 & 5$\times$5$\times$5 & & 294(2) & & 280(1) & expected\\
\hline
\hline
\end{tabular}
\end{table}


After studying finite size effects for methane hydrate using the TIP4P/Ice model, we will now replicate the simulations using the TIP4P/2005 model. While the TIP4P/2005 force field provides results that deviate more from experimental values (with a melting point of 250~K \cite{conde2017high,blazquez2022melting}), our aim is to determine whether the conclusions regarding finite-size effects on determining $T_3$ for methane hydrates remain consistent with this alternative model. Additionally, this model has demonstrated high efficiency in electrolyte simulations, a crucial aspect for studying hydrates in marine conditions \cite{BLAZQUEZ2023122031}. The configurations employed are identical to those in the previous TIP4P/Ice case. 

\begin{figure*}
     \centering
     \begin{subfigure}[hbt]{0.30\textwidth}
         \centering
         \includegraphics[width=\textwidth]{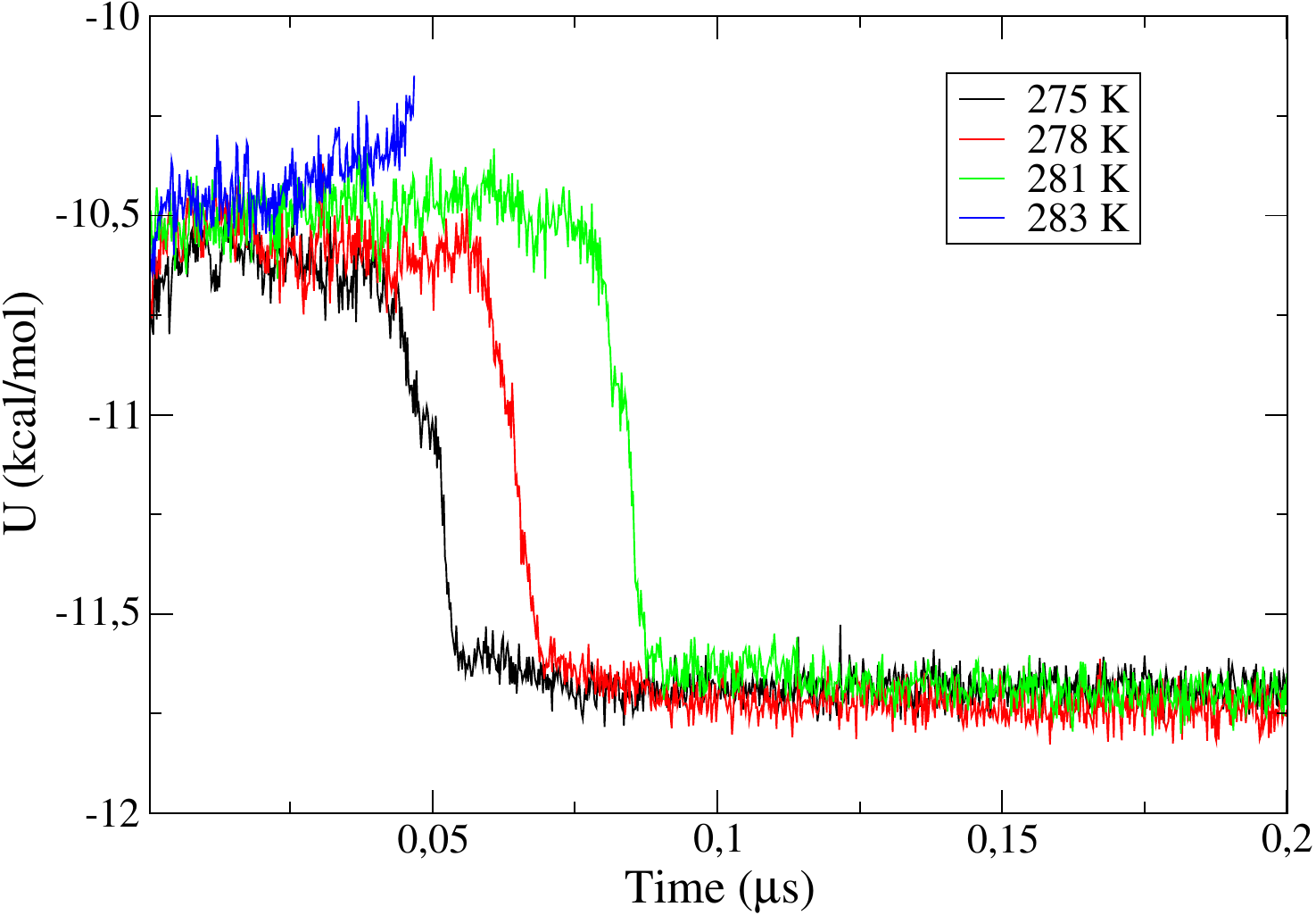}
         \caption{Conf. 1}
         \label{conf1-2005}
     \end{subfigure}
     \hfill
     \begin{subfigure}[hbt]{0.30\textwidth}
         \centering
         \includegraphics[width=\textwidth]{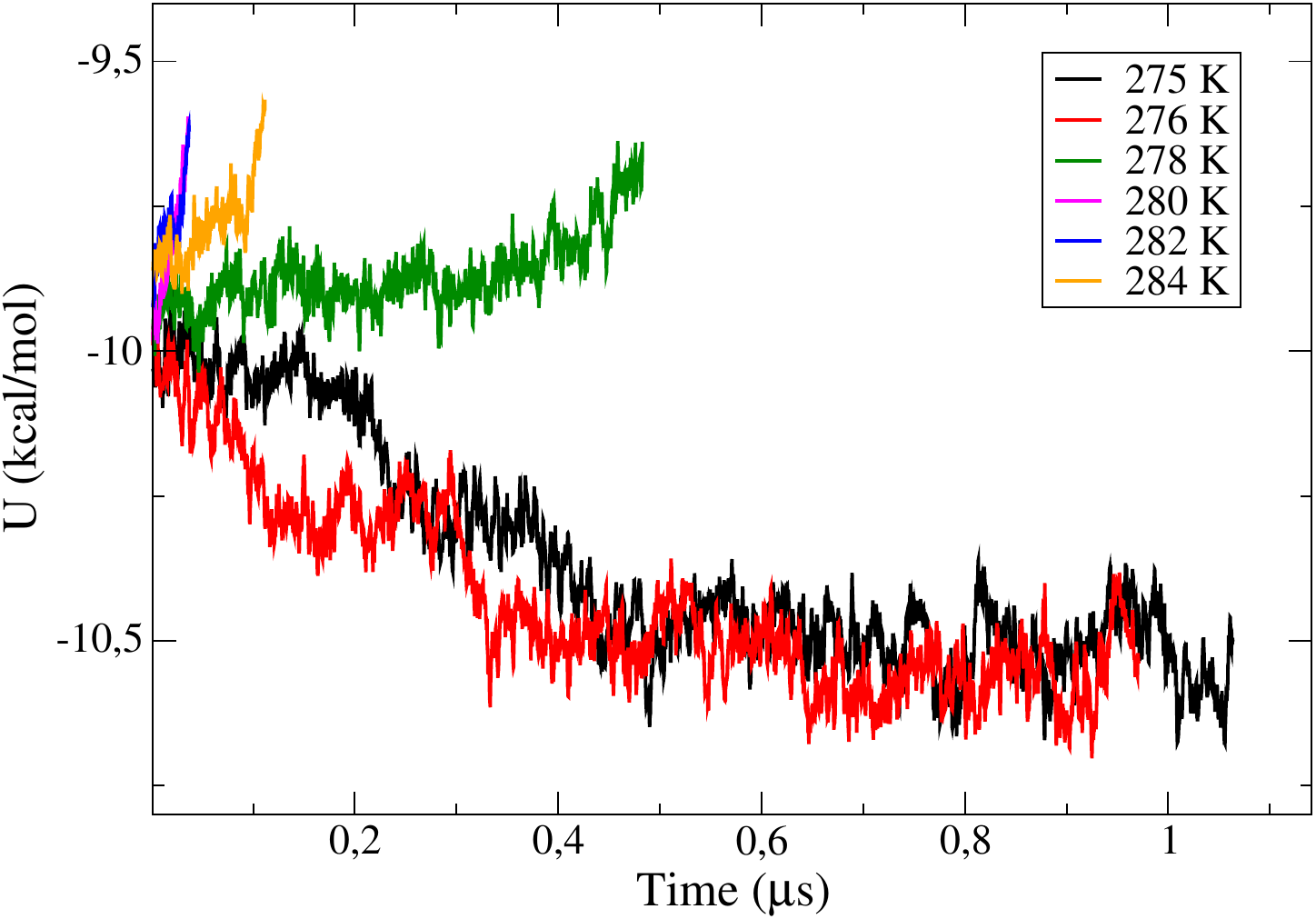}
         \caption{Conf. 2}
         \label{conf2-2005}
     \end{subfigure}
     \hfill
     \begin{subfigure}[hbt]{0.30\textwidth}
         \centering
         \includegraphics[width=\textwidth]{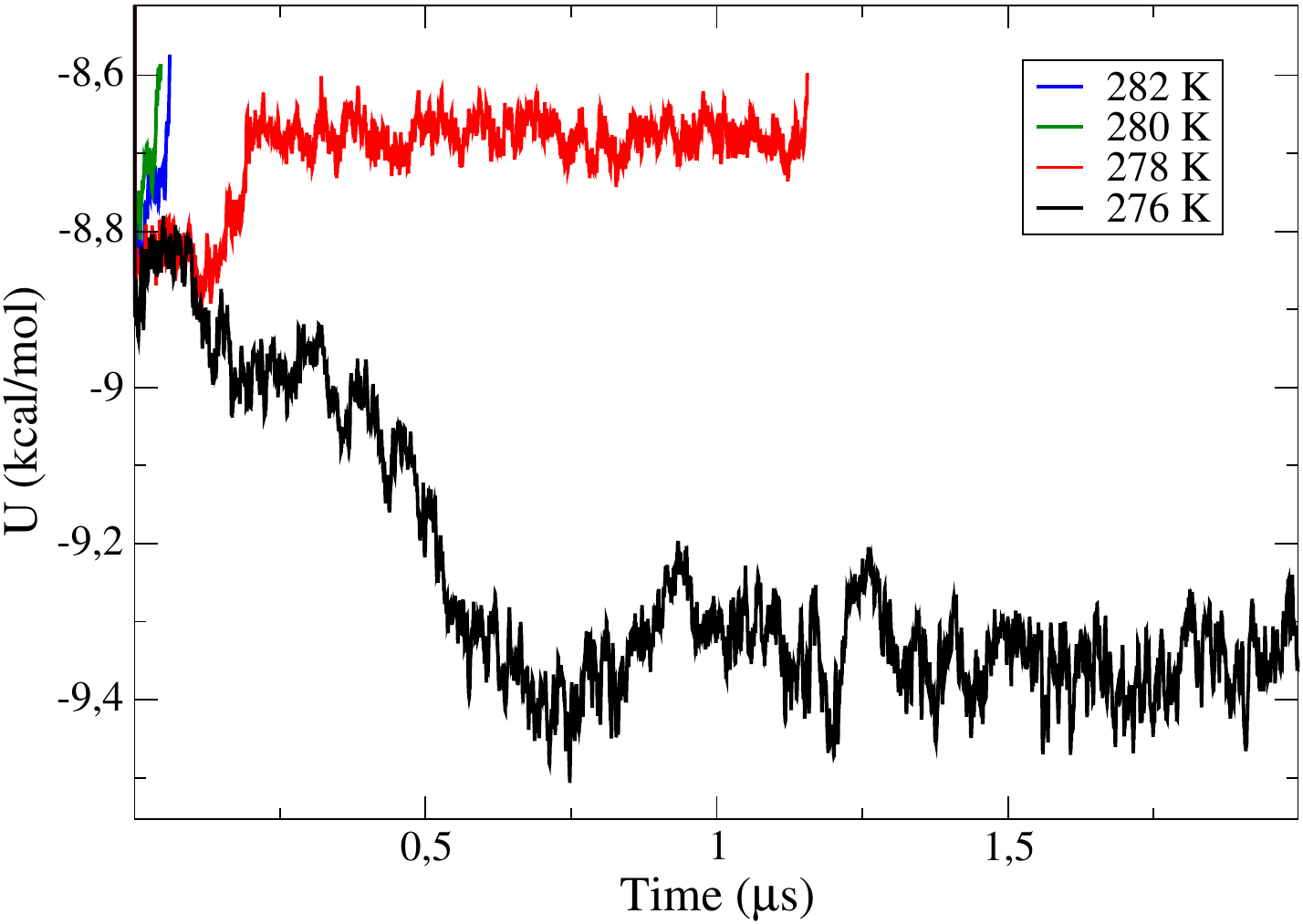}
         \caption{Conf. 3}
         \label{conf3-2005}
     \end{subfigure}
     \hfill
     \begin{subfigure}[hbt]{0.30\textwidth}
         \centering
         \includegraphics[width=\textwidth]{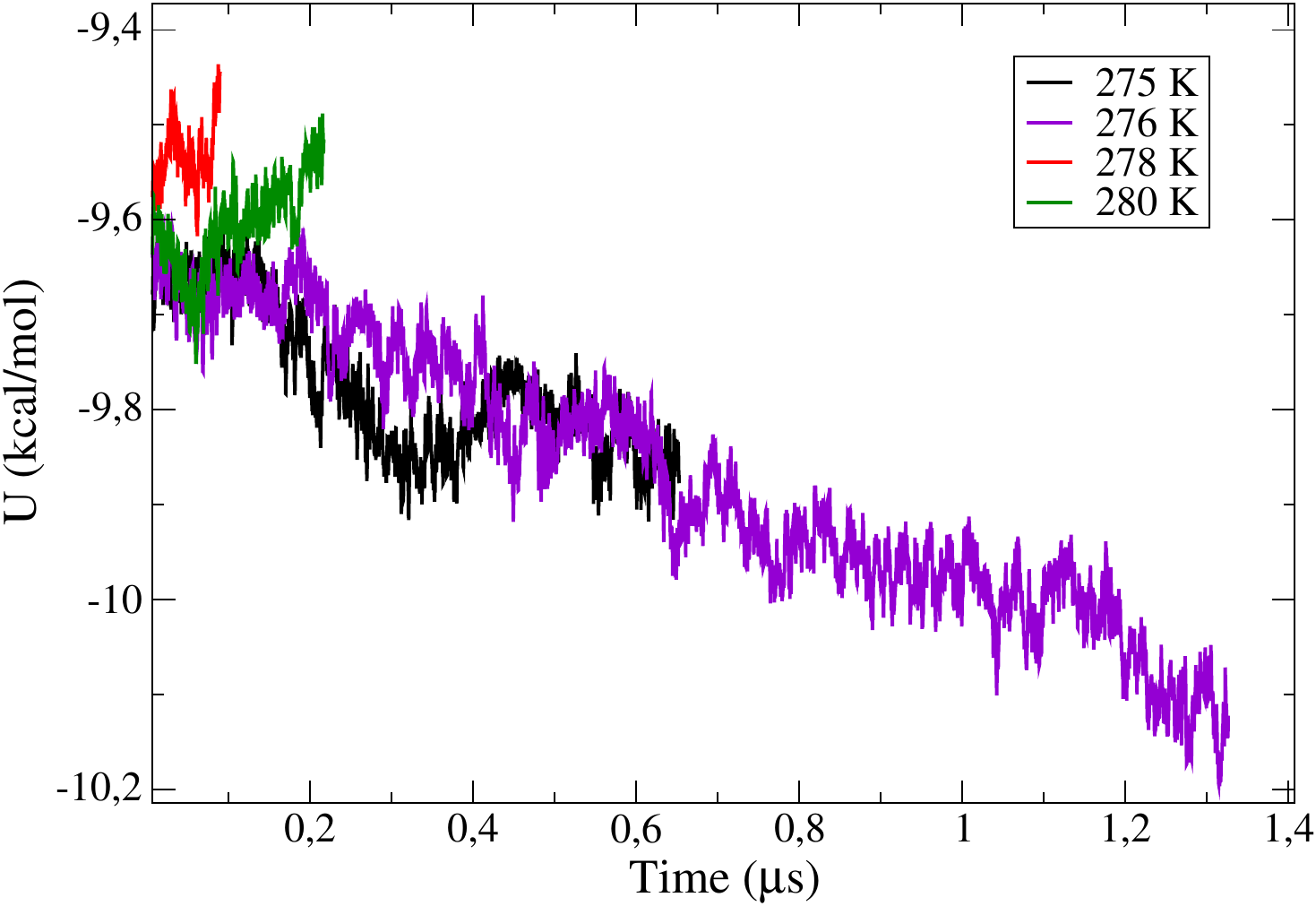}
         \caption{Conf. 4}
         \label{conf4-2005}
     \end{subfigure}
     \hfill
     \begin{subfigure}[hbt]{0.30\textwidth}
         \centering
         \includegraphics[width=\textwidth]{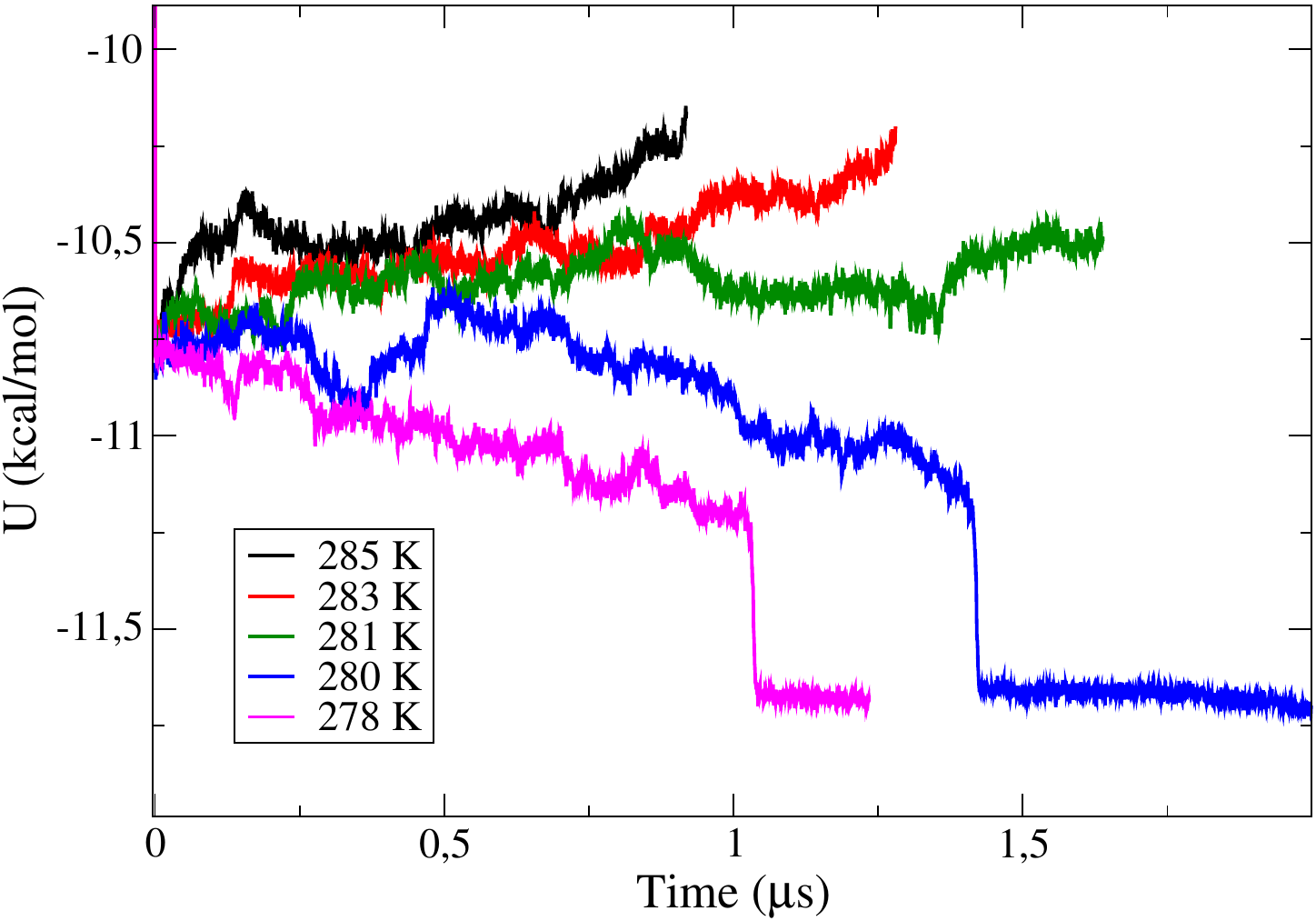}
         \caption{Conf. 5}
         \label{conf5-2005}
     \end{subfigure}
     \hfill
     \begin{subfigure}[hbt]{0.30\textwidth}
         \centering
         \includegraphics[width=\textwidth]{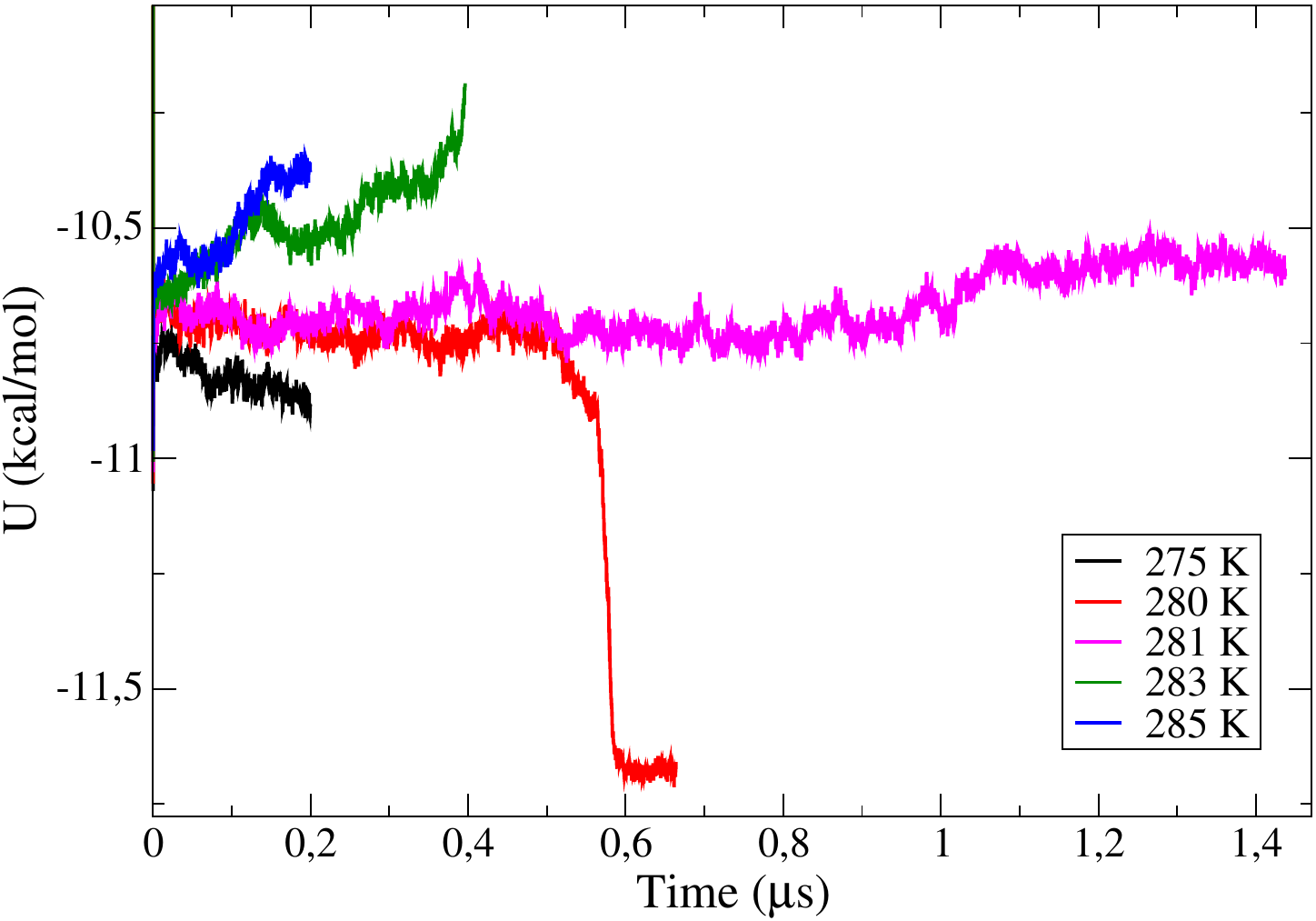}
         \caption{Conf. 6}
         \label{conf6-2005}
     \end{subfigure}
     \hfill
      \begin{subfigure}[hbt]{0.30\textwidth}
         \centering
         \includegraphics[width=\textwidth]{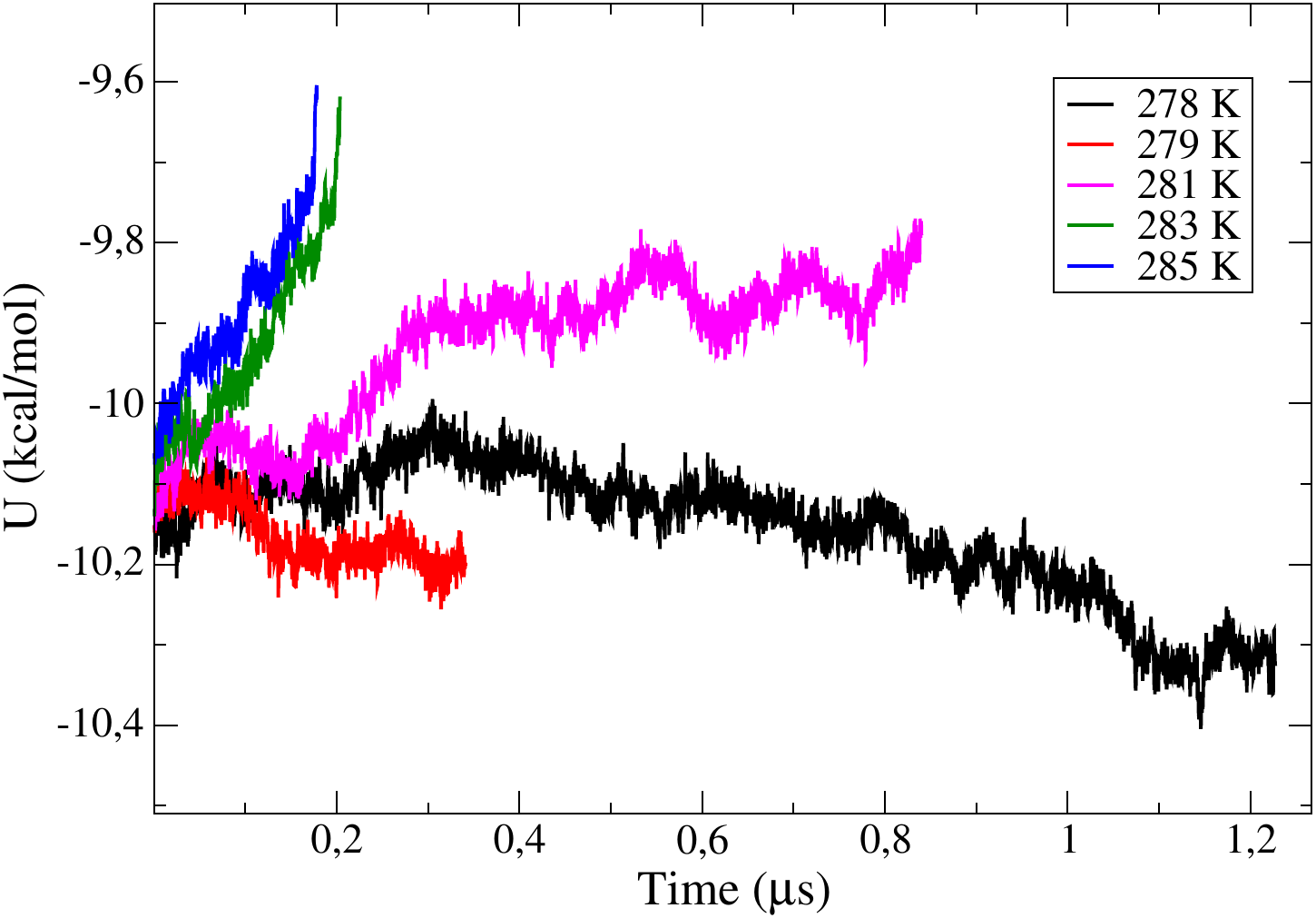}
         \caption{Conf. 7}
         \label{conf7-2005}
     \end{subfigure}
     \hfill
     \begin{subfigure}[hbt]{0.30\textwidth}
         \centering
         \includegraphics[width=\textwidth]{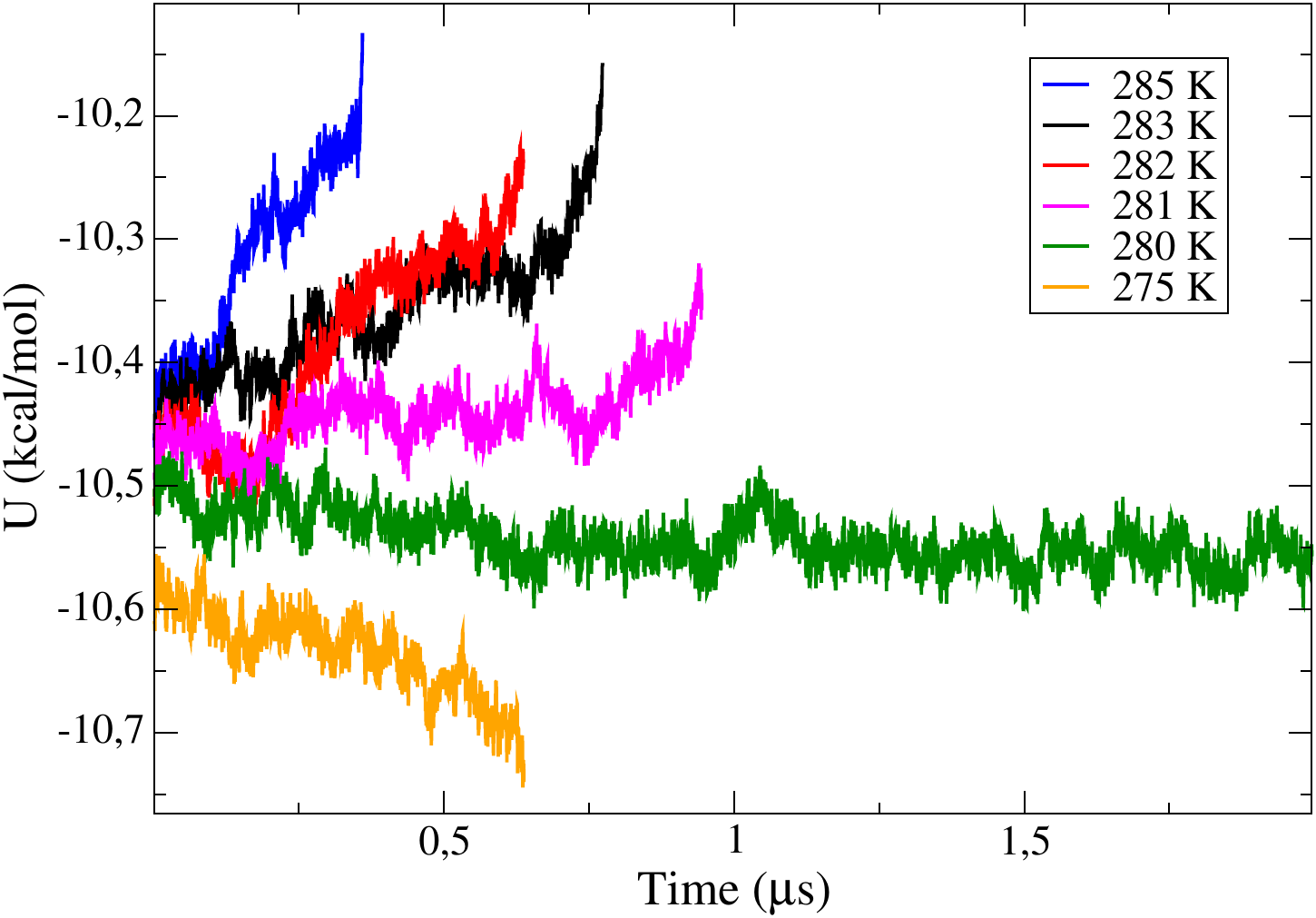}
         \caption{Conf. 8}
         \label{conf8-2005}
     \end{subfigure}
     \hfill
      \begin{subfigure}[hbt]{0.30\textwidth}
         \centering
         \includegraphics[width=\textwidth]{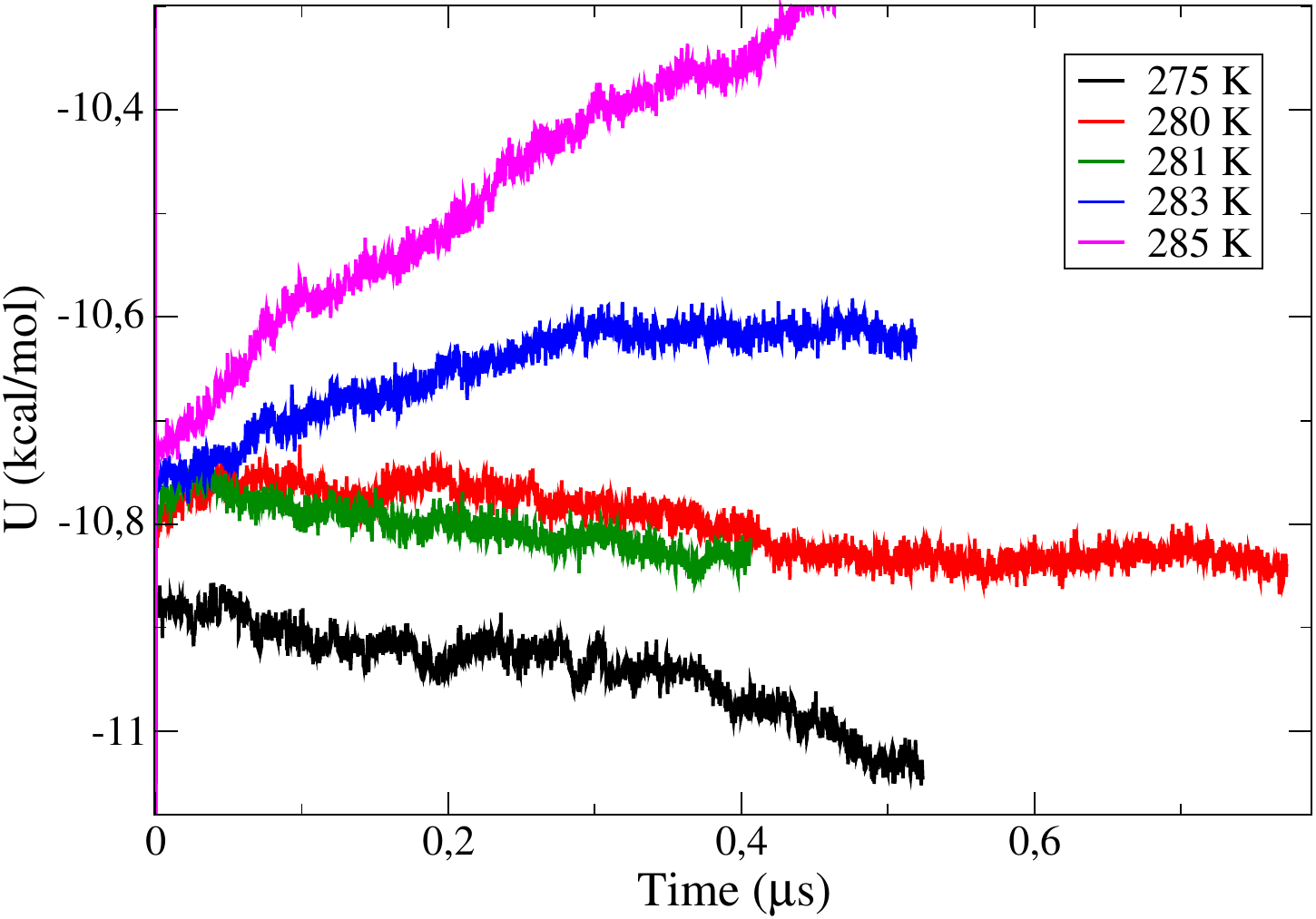}
         \caption{Conf. 9}
         \label{conf9-2005}
     \end{subfigure}
        \caption{\justifying{Evolution of the potential energy over time for the 9 size-dependent configurations analyzed in this work. The results were obtained from $NpT$ simulations at 400~bar using the TIP4P/2005 model.}}
        \label{confs-2005}
\end{figure*}

For configuration 1, similar to the TIP4P/Ice, the formation of a bubble is observed (see the abrupt drop in potential energy in Fig.~\ref{conf1-2005}). The estimated $T_3$ for this configuration is 282(1)~K, in agreement with the value published in 2010 \cite{JCP_2010_133_064507}, but deviating significantly from the experimental triple-point temperature due to the correlation with the melting point of the model \cite{conde2013note}, as previously emphasized.

Regarding configurations 2, 3, and 4, as shown in Figs.~\ref{conf2-2005}, \ref{conf3-2005}, and \ref{conf4-2005}, the values of $T_3$ remain consistent across all these configurations, estimated at 277(1)~K. This behavior is consistent with the observations in the TIP4P/Ice model. Notably, configuration 1, due to its size and the emergence of a bubble, exceeds the temperatures of these configurations by 5~K.

Next, we present the results for configurations 5 and 6, where the appearance of a bubble is observed in both cases due to the stoichiometric condition of the system, reflected by the drop in potential energy shown in Figs.~\ref{conf5-2005} and \ref{conf6-2005}, resulting in a $T_3$ of 280.5(1)~K. Finally, configurations 7, 8, and 9, regardless of whether bubble formation is expected or not, converge to the same $T_3$ value of 280(1) K. In a recent study on methane hydrates in equilibrium with aqueous NaCl solutions, Blazquez et al. \cite{BLAZQUEZ2023122031} reported a $T_3$ value of 279(1)~K for the TIP4P/2005 model in perfect agreement with the result obtained in our work for the same model.

\begin{figure*}[htp]
    \centering
    \includegraphics[width=0.45\textwidth]{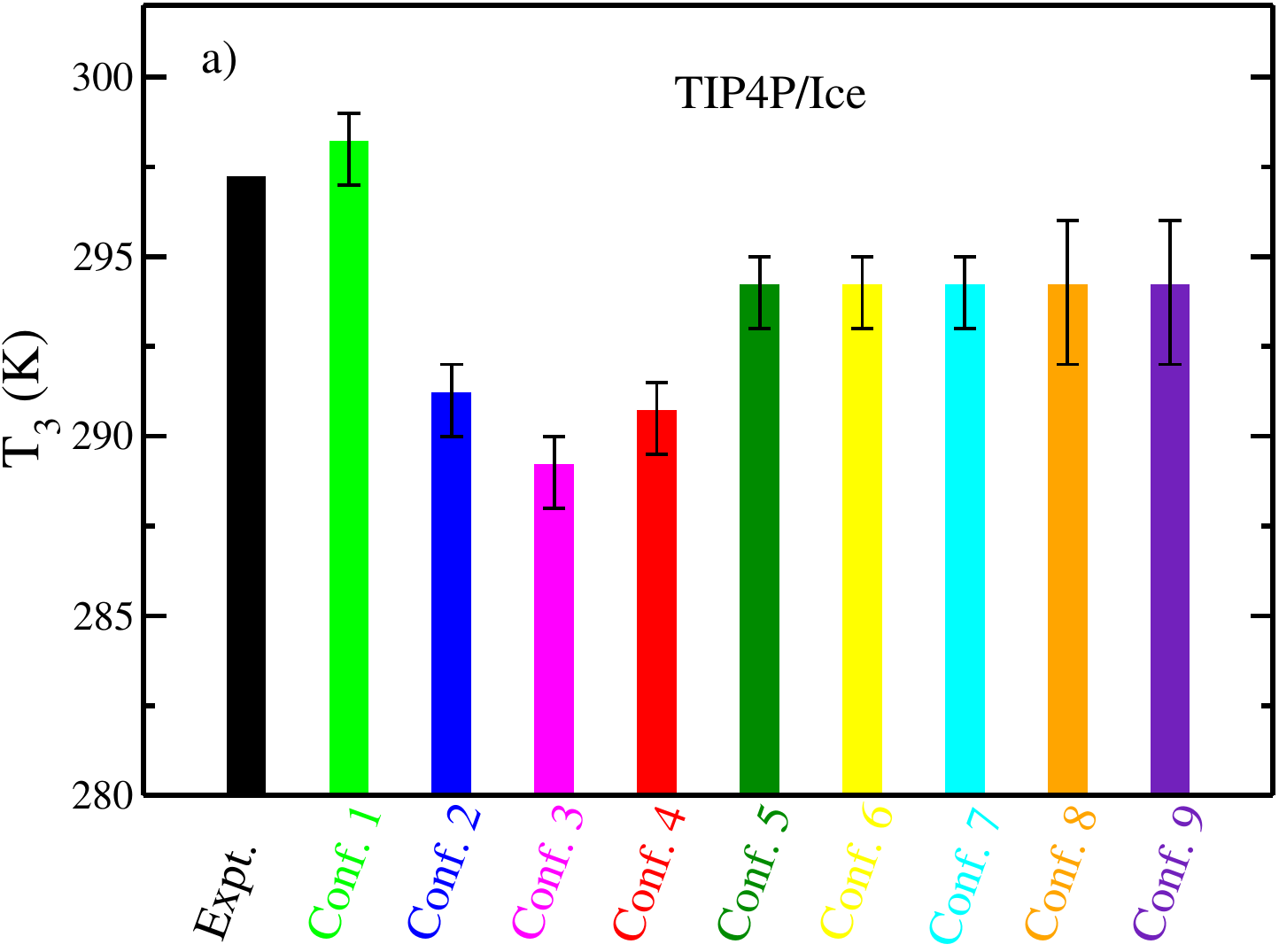}
        \includegraphics[width=0.45\textwidth]{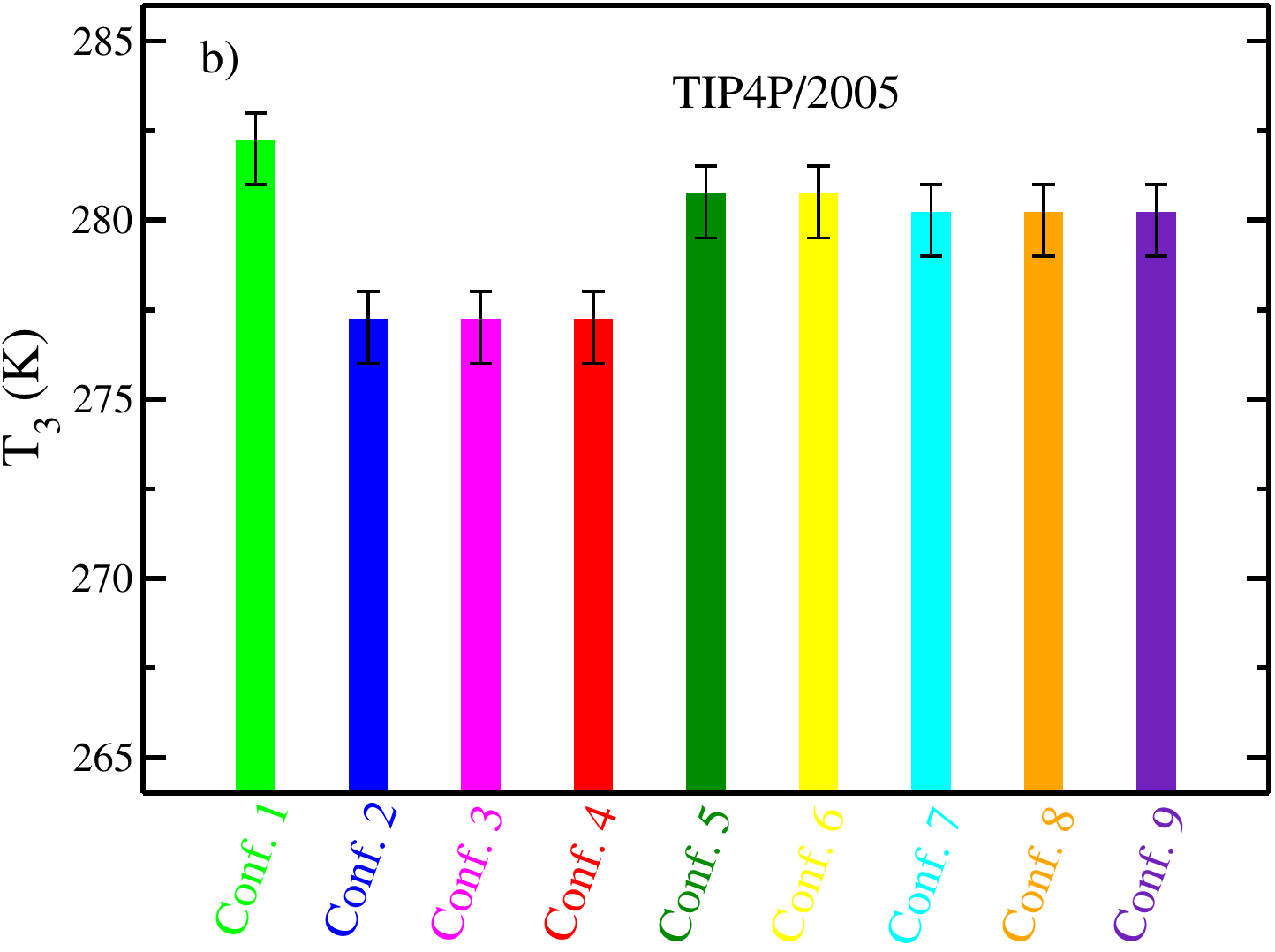}
    \caption{\justifying{
    Comparison of the three-phase coexistence temperatures ($T_3$) for methane hydrate from the 9 size-dependent configurations analyzed in this work using a) TIP4P/Ice model and b) TIP4P/2005. All simulations were performed at 400 bar. The experimental $T_3$ value is included for the TIP4P/Ice model taken from \cite{Sloan_book_hydrates}. In all cases, the composition of the initial configuration was analogous, formed by a methane hydrate phase in coexistence with a liquid water phase and a methane gas phase with the corresponding numbers of molecules in each configuration.
    }}
    \label{summary-t3}
\end{figure*}

We have summarized the $T_3$ values obtained for each of the 9 size-dependent configurations in Table~\ref{tabla-t3-todos} for both the TIP4P/Ice and TIP4P/2005 models. Overall, these models consistently replicate the same trend regarding finite-size effects in small systems. Moreover, the stoichiometric composition condition is fulfilled to observe bubble formation. Additionally, Fig.~\ref{summary-t3} presents bar graphs illustrating the $T_3$ values for each configuration. For both models, opting for configuration 1, characterized by stoichiometric composition and bubble formation, leads to an overestimation of results, indicating a false stability of the methane hydrate phase. Subsequently, a group of configurations (2, 3, and 4) exhibits $T_3$ values slightly influenced by finite-size effects. Finally, from the size of configuration 5 onwards, the $T_3$ value converges within statistical uncertainty, independently of its composition. This suggests that this is the optimal $T_3$ value in this system for each model, ensuring that finite-size effects can be safely neglected. Note that, for the TIP4P/Ice model, we have included the experimental value of $T_3$ due to its agreement with the simulation-derived value for this model. One can compare also with theoretical results, for example with those using the SAFT-VR approach in which it is demonstrated that these theoretical findings are in close agreement with experimental data\cite{dufal2012modelling} and thus, with our calculations using the TIP4P/Ice.

\section{Concluding remarks}

In this study, we have explored the finite-size effects on determining the three-phase coexistence temperature ($T_3$) for methane hydrate using Molecular Dynamics simulations with the direct coexistence technique. Specifically, we used 9 size-dependent configurations and two realistic water models (i.e., TIP4P/Ice and TIP4P/2005) to evaluate the impact of size and composition on the estimation of $T_3$. Our findings, valid for both models, can be summarized as follows:

\begin{itemize}
    \item {The study confirms the sensitivity of $T_3$ depending on the size and composition of the system, explaining the discrepancies observed in the original work by Conde and Vega in 2010.}
    \item {Configurations with stoichiometric composition or with less gas molecules than those required by the stoichiometric composition of the hydrate, at temperatures below $T_3$ evolve into a singular phase of methane hydrate growth, characterized by the emergence of a bubble within the liquid and the subsequent formation of an oversaturated methane solution in water. Conversely, an excess of molecules in the gas phase in the initial configuration leads to the coexistence of methane hydrate and methane gas phases without the formation of bubbles.}
    \item {Finite-size effects were pronounced in small systems with stoichiometric composition (e.g., configuration 1 with a unit cell of $2\times2\times2$), resulting in an overestimation of $T_3$ due to bubble formation during hydrate growth, causing a false stability of methane hydrate by increasing methane solubility.}
    \item {Configurations with larger unit cells (e.g., $3\times3\times3$ and beyond) show convergence of $T_3$ values, suggesting that finite-size effects for these system sizes, regardless of bubble formation, can be safely neglected.}
    \item {The study emphasizes the significant impact of methane bubble formation in small-sized systems with stoichiometric composition, influencing stability and overestimation of $T_3$ values. In larger stoichiometric systems, bubble formation accelerates complete hydrate growth but does not affect the $T_3$ value, as long as $T_3$ is determined before the potential energy drop (i.e., before the bubble formation).}
    \item {To study the $T_3$ of methane hydrate the best choice is configuration 7, which provides an accurate $T_3$ value, and  affordable simulation times.
    For highly soluble gases (as CO$_2$) we also recommend to use configuration 7 but increasing the number of initial molecules in the gas phase (for instance from 400 to 800 to account for the much higher solubility of carbon dioxide). }
    \item {For the $T_3$ of methane hydrate at 400 bar and using the TIP4P/Ice force field we have reached a consensus: The $T_3$ is 294(2)~K. For the TIP4P/2005 model, the consensus on $T_3$ is 280(2)~K. These values could be used as a benchmark for testing new methodologies in the future.
}
\end{itemize}

In summary, considering the results, the message is clear: to estimate $T_3$ in methane hydrate systems using the direct coexistence technique, one must avoid small stoichiometric configurations like configuration 1, which form bubbles at the beginning of the run and overestimate the $T_3$ value. For larger stoichiometric configurations like 5 and 6 onwards, it is essential to proceed with caution. Ensuring that $T_3$ is determined before the abrupt energy drop is crucial. Most of the time, it will coincide, but one must ensure not to determine $T_3$ when the stability of the system is simulated erroneously. Undoubtedly, the best choice is configuration 7, providing an accurate $T_3$ value, and simulation times for this system are computationally accessible today, making configuration 7 the ideal compromise between time and size. 

This work provides valuable information on the size-dependent behavior of methane hydrate systems and offers practical recommendations to avoid finite-size effects in $T_3$ estimation through careful selection of system configurations. We anticipate that our findings will contribute to understanding finite size effects in determining $T_3$ for methane hydrates, addressing discrepancies found in the literature and aiding researchers in choosing appropriate system sizes for future studies. Our future steps will focus on studying the potential impact of cutoff values and guest types on $T_3$ values and how they are influenced by finite size effects. 


\section*{Supplementary material}
See the supplementary material for the movie of the simulation trajectory at $290\,\text{K}$ and $400\,\text{bar}$ for configuration $6$ using the TIP4P/Ice model. The movie illustrates the diffusion of methane molecules from the gas phase to the liquid phase and the formation of the bubble. 


\section*{Acknowledgments}
This work was funded by Grants No. PID2019-105898GB-C21, PID2019-105898GA-C22, PID2022-136919NB-C31 and PID2022-136919NB-C32 of the MICINN and by Project No. ETSII-UPM20-PU01 from ``Ayudas Primeros Proyectos de la ETSII-UPM''. M.M.C. acknowledges CAM and UPM for financial support of this work through the CavItieS project No. APOYO-JOVENES-01HQ1S-129-B5E4MM from ``Accion financiada por la Comunidad de Madrid en el marco del Convenio Plurianual con la Universidad Politecnica de Madrid en la linea de actuacion estimulo a la investigacion de jovenes doctores'' and CAM under the Multiannual Agreement with UPM in the line Excellence Programme for University Professors, in the context of the V PRICIT (Regional Programme of Research and Technological Innovation). This work was also funded by Ministerio de Ciencia e Innovaci\'on (Grant No.~PID2021-125081NB-I00), Junta de Andalucía (P20-00363), and Universidad de Huelva (P.O. FEDER UHU-1255522 and FEDER-UHU-202034), all four cofinanced by EU FEDER funds. The authors gratefully acknowledge the Universidad Politecnica de Madrid (www.upm.es) for providing computing resources on Magerit Supercomputer. S.B. acknowledges Ayuntamiento de Madrid for a Residencia de Estudiantes grant.\\

\section*{AUTHORS DECLARATIONS}

\section*{Conflicts of interest}

The authors have no conflicts to disclose.

\section*{Data availability}

The data that support the findings of this study are available within the article.

\section*{REFERENCES}
\bibliographystyle{ieeetr}
\bibliography{bibliografia,aipsamp}
\end{document}